\documentclass[aps,prd,superscriptaddress,twocolumn]{revtex4-1}
\usepackage{amssymb}
\usepackage{amsmath}
\usepackage{wallpaper}
\usepackage{bm}
\usepackage{multirow}
\usepackage[colorlinks,linkcolor=blue, urlcolor=blue, anchorcolor=blue, citecolor=blue]{hyperref}

\begin{document}

\title{Unified treatment for in-medium light and heavy clusters with RMF models}

\author{Cheng-Jun Xia}
\email{cjxia@yzu.edu.cn}
\affiliation{Center for Gravitation and Cosmology, College of Physical Science and Technology, Yangzhou University, Yangzhou 225009, China}

\date{\today}

\begin{abstract}
It was shown that light nuclei such as $^4$He, $^8$Be, and $^{12}$C can be well described by RMF models, which enables a unified description for nuclei with baryon numbers $A\gtrsim4$. In this work, we propose a hybrid treatment for investigating the clustering phenomenon in nuclear medium, where clusters ranging from light nuclei (e.g., $^3$H, $^3$He, and $^4$He) to heavy ones (e.g., $^{12}$C, $^{16}$O, $^{40}$Ca, $^{48}$Ca, and $^{208}$Pb) can be treated in a unified manner. In particular, assuming a spherical Wigner-Seitz cell, the clusters are fixed by solving the Dirac equations imposing the Dirichlet-Neumann boundary condition, while the nuclear medium are treated with Thomas-Fermi approximation and take constant densities. In the presence of nuclear medium, the clusters eventually become unbound as density increases, while the root-mean-square charge radii increase. For clusters with different proton and neutron numbers $N_p \neq N_n$, their binding energies varies with the proton fraction of nuclear medium, which are less significant for clusters with $N_p = N_n$. The uncertainties of density functionals on the clustering phenomenon are investigated as well adopting 8 different functionals. Based on the obtained results, an analytical formula describing the binding energies of in-medium clusters is then obtained. The results presented in this work should be useful to understand the clustering phenomenon in both heavy-ion collisions and neutron stars.
\end{abstract}

\maketitle

\section{\label{sec:intro}Introduction}

At sufficiently small densities, the clustering phenomenon is expected to take place in nuclear medium and produces various nuclei, which affects the equation of state (EOS), composition, and various transport properties of low density nuclear matter~\cite{Roepke2015_arXiv1501-01222, Lu2020_PRL125-192502}. In particular, the clustering will take place and play an important role in various scenarios such as the preformation of $\alpha$ in finite nuclei~\cite{Ebran2012_Nature487-341, Roepke2014_PRC90-034304, Xu2016_PRC93-011306, Tohsaki2017_RMP89-011002}, crusts of neutron stars~\cite{Xia2022_CTP74-095303, Pais2025}, heavy-ion collisions~\cite{Qin2012_PRL108-172701, Bougault2020_JPG47-025103, Custodio2025_PRL134-082304}, core-collapse supernovae~\cite{Arcones2008_PRC78-015806, Fischer2020_PRC102-055807}, and binary neutron star mergers~\cite{Rosswog2015_IJMPD24-1530012, Alford2018_PRL120-041101, Fujibayashi2018_ApJ860-64}. It is thus essential to unveil the clustering phenomenon in nuclear matter under various conditions.

As commonly adopted in nuclear astrophysics, the nuclear statistical equilibrium (NSE) models can be employed to describe the clustering phenomenon~\cite{Meyer1994_ARAA32-153}, where nuclear matter is treated as a system comprised of non/minimally-interacting nuclei with a distribution of species determined from nuclear masses according to statistical equilibrium. Nevertheless, as density increases, the interaction between nuclei should not be neglected~\cite{Roepke1982_NPA379-536, Horowitz2006_NPA776-55}, where Mott transition eventually takes place with clusters become unbound due to Pauli blocking. To accommodate cluster melting at larger densities, an excluded volume mechanism for light clusters was latter introduced~\cite{Lattimer1991_NPA535-331, Shen2011_ApJ197-20}, which improves the NSE models~\cite{Hempel2010_NPA837-210, Raduta2010_PRC82-065801}. Meanwhile, in the framework of relativistic-mean-field (RMF) models, a generalized relativistic density-functional (gRDF) approach was proposed to treat the clustering phenomenon in nuclear medium~\cite{Typel2010_PRC81-015803}, where light clusters are considered as quasi-particles with the binding energies varying with density and temperature. The binding energy shifts of clusters are fixed within a quantum statistical approach~\cite{Roepke1982_NPA379-536, Roepke2014_JP569-012031, Roepke2020_PRC101-064310}, while the strong interactions among clusters and nucleons are mediated via meson exchange~\cite{Typel2010_PRC81-015803, Avancini2010_PRC82-025808, Ferreira2012_PRC85-055811, Fischer2020_PRC102-055807}.

Despite their great successes, additional parametrizations were introduced in both the excluded volume and gRDF approaches, while the clusters considered are restricted to light nuclei with baryon number $A\leq 16$~\cite{Typel2010_PRC81-015803, Avancini2010_PRC82-025808, Ferreira2012_PRC85-055811, Fischer2020_PRC102-055807, Roepke2020_PRC101-064310}. A unified treatment for in-medium light and heavy clusters with well calibrated parameters reproducing finite nuclei properties are thus favorable. In particularly, relativistic density functionals typically have deep single-nucleon potentials, which predict the occurrence of much more pronounced cluster structures than non-relativistic density functionals~\cite{Ebran2012_Nature487-341}. Since the RMF models give satisfactory description for both finite nuclei~\cite{Brockmann1977_PLB69-167, Boguta1981_PLB102-93, Mares1989_ZPA333-209, Reinhard1989_RPP52-439, Toki1994_PTP92-803, Ring1996_PPNP37_193-263, Meng2006_PPNP57-470, Paar2007_RPP70-691, Tanimura2012_PRC85-014306, Wang2013_CTP60-479, Meng2015_JPG42-093101, Meng2016_RDFNS, Chen2021_SCPMA64-282011, Typel1999_NPA656-331, Vretenar1998_PRC57-R1060, Lu2011_PRC84-014328, Wei2020_CPC44-074107, Taninah2020_PLB800-135065} and nuclear matter~\cite{Glendenning2000, Ban2004_PRC69-045805, Weber2007_PPNP59-94, Long2012_PRC85-025806, Sun2012_PRC86-014305, Wang2014_PRC90-055801, Fedoseew2015_PRC91-034307, Gao2017_ApJ849-19}, in this work we adopt RMF models to investigate the clustering phenomenon in nuclear medium.


To this end, we propose a hybrid treatment for investigating the clustering phenomenon in nuclear medium within a spherical Wigner-Seitz cell. The clusters are fixed by solving the Dirac equations imposing the Dirichlet-Neumann boundary condition~\cite{Negele1973_NPA207-298}, while the nuclear medium are treated with Thomas-Fermi approximation and take constant densities. The microscopic treatment including both direct and exchange terms for the center-of-mass corrections of clusters are adopted~\cite{Bender2000_EPJA7-467}, which was shown to play an important role in properly describing the binding energies of light nuclei~\cite{Rong2023_PRC108-054314}. As the density of nuclear medium increases, it is found that the clusters eventually become unbound with negative binding energies while the root-mean-square charge radii $R_\mathrm{ch}$ increase. The binding energy shifts of clusters with respect to the proton fraction of nuclear medium are examined as well.

The paper is organized as follows. In Sec.~\ref{sec:the} we present our theoretical framework, including the Lagrangian density of RMF models and formalism for fixing the properties of clusters in nuclear medium. The obtained results are presented in Sec.~\ref{sec:res}.  We draw our conclusion in Sec.~\ref{sec:con}.

\section{\label{sec:the}Theoretical framework}
\subsection{\label{sec:the_Lagrangian} Lagrangian density}
The Lagrangian density of RMF models can be obtained with
\begin{eqnarray}
\mathcal{L}
 &=& \sum_{i=n,p} \bar{\psi}_i
       \left[  i \gamma^\mu \partial_\mu - \gamma^0 \left(g_\omega\omega + g_\rho\rho\tau_i + A q_i\right)- m_i^* \right] \psi_i
\nonumber \\
 &&\mbox{} + \frac{1}{2}\partial_\mu \sigma \partial^\mu \sigma  - \frac{1}{2}m_\sigma^2 \sigma^2
           - \frac{1}{4} \omega_{\mu\nu}\omega^{\mu\nu} + \frac{1}{2}m_\omega^2 \omega^2
\nonumber \\
 &&\mbox{} - \frac{1}{4} \rho_{\mu\nu}\rho^{\mu\nu} + \frac{1}{2}m_\rho^2 \rho^2 - \frac{1}{4} A_{\mu\nu}A^{\mu\nu} + U(\sigma, \omega),
\label{eq:Lagrange}
\end{eqnarray}
where $\psi_{i}$ is the Dirac spinor, $m_{n,p}^*\equiv m_{n,p} + g_{\sigma} \sigma$ the effective nucleon mass, $\tau_n=-\tau_p=1$ the 3rd component of isospin and $q_i$ the charge number of fermion $i$ with $q_{n}=0$ and $q_{p}=-q_{e}=1$. The boson fields $\sigma$, $\omega$, $\rho$, and $A$ take mean values and are left with only the time components due to time-reversal symmetry, so that the field tensors $\omega_{\mu\nu}$, $\rho_{\mu\nu}$, and $A_{\mu\nu}$ vanish except for
\begin{equation}
\omega_{i0} = -\omega_{0i} = \partial_i \omega,
 \rho_{i0}  = -\rho_{0i}   = \partial_i  \rho,
  A_{i0}    = -A_{0i}      = \partial_i A. \nonumber
\end{equation}

The mesonic nonlinear self couplings in Eq.~(\ref{eq:Lagrange}) are fixed with
\begin{equation}
U(\sigma, \omega) = -\frac{1}{3}g_2\sigma^3 - \frac{1}{4}g_3\sigma^4 + \frac{1}{4}c_3\omega^4,  \label{eq:U_NL}
\end{equation}
which are adopted for the relativistic density functionals NL3~\cite{Lalazissis1997_PRC55-540}, PK1~\cite{Long2004_PRC69-034319}, and TM1~\cite{Sugahara1994_NPA579-557}.
Alternatively, we could employ the density-dependent couplings with
\begin{eqnarray}
g_{\xi}(n_\mathrm{b}) &=& g_{\xi} a_{\xi} \frac{1+b_{\xi}(n_\mathrm{b}/n_0+d_{\xi})^2}
                          {1+c_{\xi}(n_\mathrm{b}/n_0+d_{\xi})^2}, \label{eq:ddcp_TW} \\
g_{\rho}(n_\mathrm{b}) &=& g_{\rho} \exp{\left[-a_\rho(n_\mathrm{b}/n_0 + b_\rho)\right]}, \label{eq:ddcp_rho}
\end{eqnarray}
where $\xi=\sigma$, $\omega$ and the baryon number density $n_\mathrm{b} = n_p+n_n$ with $n_0$ being the saturation density. This formalism is adopted for the relativistic density functionals PKDD~\cite{Long2004_PRC69-034319}, DD2~\cite{Typel2010_PRC81-015803}, DD-ME2~\cite{Lalazissis2005_PRC71-024312}, DD-MEX~\cite{Taninah2020_PLB800-135065}, and DD-LZ1~\cite{Wei2020_CPC44-074107}.

\subsection{\label{sec:the_Dirac} Clusters in nuclear medium}
To fix the properties of clusters in nuclear medium, we assume they are spherically symmetric so that the Dirac spinors of nucleons can be expanded as
\begin{equation}
 \psi_{n\kappa m}({\bm r}) =\frac{1}{r}
 \left(\begin{array}{c}
   iG_{n\kappa}(r) \\
    F_{n\kappa}(r) {\bm\sigma}\cdot{\hat{\bm r}} \\
 \end{array}\right) Y_{jm}^l(\theta,\phi)\:,
\label{EQ:RWF}
\end{equation}
with $G_{n\kappa}(r)/r$ and $F_{n\kappa}(r)/r$ being the radial wave functions for the upper and lower components, while $Y_{jm}^l(\theta,\phi)$
is the spinor spherical harmonics. The quantum number $\kappa$ is connected to the angular momenta $(l,j)$ via $\kappa=(-1)^{j+l+1/2}(j+1/2)$. The Dirac equation for the radial wave functions is then
\begin{equation}
 \left(\begin{array}{cc}
  V_i+S_i                             & {\displaystyle -\frac{\mbox{d}}{\mbox{d}r}+\frac{\kappa}{r}}\\
  {\displaystyle \frac{\mbox{d}}{\mbox{d}r}+\frac{\kappa}{r}} & V_i-S_i-2m_i                       \\
 \end{array}\right)
 \left(\begin{array}{c}
  G_{n\kappa} \\
  F_{n\kappa} \\
 \end{array}\right)
 = \varepsilon_{n\kappa}
 \left(\begin{array}{c}
  G_{n\kappa} \\
  F_{n\kappa} \\
 \end{array}\right) \:,
\label{EQ:RDirac}
\end{equation}
where $\varepsilon_{n\kappa}$ represents the single nucleon energy with the scalar ($S_i$) and vector ($V_i$) potentials fixed by
\begin{eqnarray}
S_i &=&  g_{\sigma} \sigma, \label{eq:MF_scalar} \\
V_i &=&  \Sigma^\mathrm{R} + g_{\omega} \omega + g_{\rho}\tau_{i} \rho + q_i  A, \label{eq:MF_vector}
\end{eqnarray}
Note that the ``rearrangement" term $\Sigma^\mathrm{R}$ emerges if the density-dependent couplings are adopted in the Lagrangian density~\cite{Lenske1995_PLB345-355}, i.e.,
\begin{equation}
\Sigma^\mathrm{R}=
 \frac{\mbox{d} g_\sigma}{\mbox{d} n_\mathrm{b}} \sigma n_\mathrm{s}+
   \frac{\mbox{d} g_\omega}{\mbox{d} n_\mathrm{b}} \omega n_\mathrm{b}+
   \frac{\mbox{d} g_\rho}{\mbox{d} n_\mathrm{b}} \rho \sum_i\tau_i n_i.
\label{eq:re_B}
\end{equation}
The equations of motion for bosons are fixed by
\begin{eqnarray}
(-\nabla^2 + m_\sigma^2) \sigma &=& -g_{\sigma} \sum_{i=n,p}n_i^\mathrm{s} - g_2\sigma^2 - g_3\sigma^3, \label{eq:KG_sigma} \\
(-\nabla^2 + m_\omega^2) \omega &=& g_{\omega} n_\mathrm{b} + c_3\omega^3, \label{eq:KG_omega}\\
(-\nabla^2 + m_\rho^2) \rho     &=& \sum_{i=n,p} g_{\rho}\tau_{i} n_i, \label{eq:KG_rho}\\
                   -\nabla^2 A  &=& e(n_p - n_e), \label{eq:KG_photon}
\end{eqnarray}
where reflective boundary conditions are imposed at $r=0$ and the edge of Wigner-Seitz cell with $r=R_\mathrm{W}$~\cite{Xia2021_PRC103-055812}. Note that in our calculation, the global charge neutrality is imposed by adding negatively charged electron background $n_e$ so that the Coulomb potential vanishes at the boundary of the Wigner-Seitz cell, i.e., $A(R_\mathrm{W})=\nabla A(R_\mathrm{W})=0$. The source currents of fermion $i$ in Eqs.~(\ref{eq:KG_sigma}-\ref{eq:KG_photon}) are $n_{i} = \langle \bar{\psi}_{i} \gamma^{0} \psi_{i} \rangle = n_{i,\mathrm{gas}} + n_{i,\mathrm{cluster}}$ and $n_{i}^\mathrm{s} = \langle \bar{\psi}_{i} \psi_{i} \rangle=n^\mathrm{s}_{i,\mathrm{gas}} + n^\mathrm{s}_{i,\mathrm{cluster}}$, which include contributions from both the nuclear medium (gas) and cluster. Adopting no-sea approximation, the contribution of clusters can be fixed according to the radial wave functions, i.e.,
\begin{eqnarray}
 n_{i,\mathrm{cluster}}^{s}(r) &=& \frac{1}{4\pi r^2}\sum_{k=1}^{N_i}  \left[|G_{k i}(r)|^2-|F_{k i}(r)|^2\right], \label{eq:np_cluster} \\
 n_{i,\mathrm{cluster}}(r) &=& \frac{1}{4\pi r^2}\sum_{k=1}^{N_i}  \left[|G_{k i}(r)|^2+|F_{k i}(r)|^2\right],
\end{eqnarray}
where $N_i$ represents the number of nucleons in clusters. Meanwhile, adopting Thomas-Fermi approximation (TFA), the scalar and vector densities for the nuclear medium are determined by
\begin{eqnarray}
n_{i,\mathrm{gas}}^\mathrm{s}(r) &=&  \frac{{m_i^*}^3}{2\pi^2} \left[ f(x_i) - f(\bar{x}_i)  \right], \label{eq:nsgas}\\
n_{i,\mathrm{gas}}           &=&  \frac{\nu_i^3-\bar{\nu}_i^3}{3\pi^2} = \mathrm{constant}, \label{eq:ngas}
\end{eqnarray}
where $\nu_i(r)$ represents the Fermi momentum and $f(x) = x \sqrt{x^2+1} - \mathrm{arcsh}(x)$ with ${x}_i =\nu_i/m_i^*$ and $\bar{x}_i =\bar{\nu}_i/m_i^*$. Meanwhile, $\bar{\nu}_i(r)$ is the minimum momentum corresponding to the lowest energy state with vanishing momentum in nuclear medium
\begin{equation}
\varepsilon_{i, \mathrm{gas}}^\mathrm{b} = S_{i, \mathrm{gas}} + V_{i, \mathrm{gas}},
\end{equation}
which is fixed for uniform nuclear matter in the absence of clusters with $n_{i,\mathrm{cluster}}^{s}=n_{i,\mathrm{cluster}}=0$. Once $\varepsilon_{i, \mathrm{gas}}^\mathrm{b}$ is fixed, $\bar{\nu}_i(r)$ can be obtained with
\begin{equation}
\sqrt{\bar{\nu}_i^2+{m_i^*}^2} + \Sigma^\mathrm{R} + g_{\omega} \omega + g_{\rho}\tau_{i} \rho + q_i  A \equiv \varepsilon_{i, \mathrm{gas}}^\mathrm{b}.  \label{eq:nub_gas}
\end{equation}
Then the Fermi momentum $\nu_i(r)$ can be fixed according to Eq.~(\ref{eq:ngas}) with constant densities $n_{i,\mathrm{gas}}$ for nuclear medium.

\begin{figure}[!ht]
  \centering
  \includegraphics[width=\linewidth]{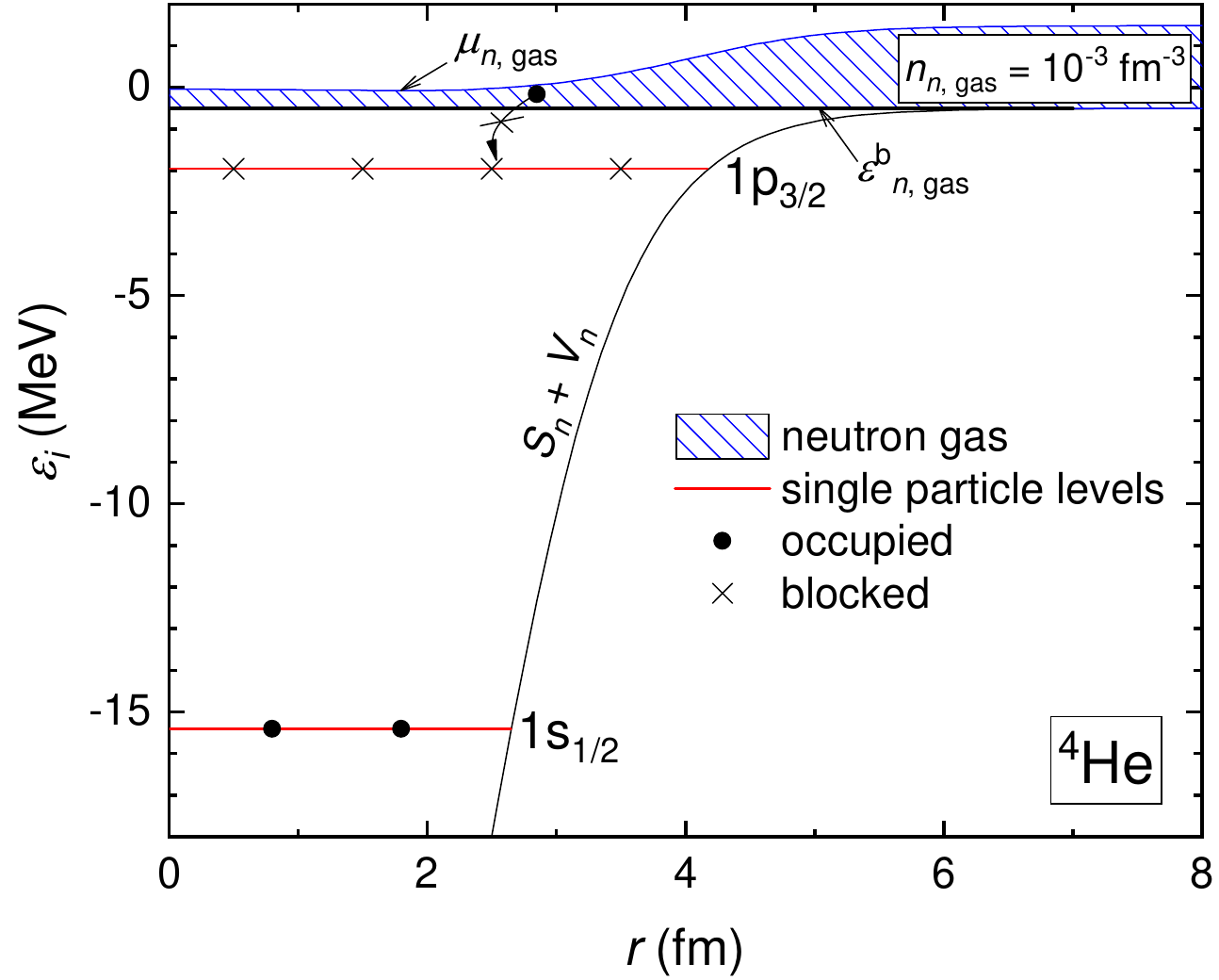}
  \caption{\label{Fig:Potential} Schematic illustration of the single neutron energy levels and their mean-field potential for a cluster $^4$He emersed in a neutron gas with $n_{n,\mathrm{gas}} = 10^{-3}\ \mathrm{fm}^{-3}$. As indicated by the solid circles, the orbital 1s$_{1/2}$ is occupied by two neutrons forming the cluster $^4$He. The lowest energy state $\varepsilon_{n, \mathrm{gas}}^\mathrm{b}$ in neutron gas gain nonzero momentum $\bar{\nu}(r)>0$ under the attractive interaction of the cluster. The energy levels in between the lowest gas state $\varepsilon_{n, \mathrm{gas}}^\mathrm{b}$ and the occupied cluster state $\varepsilon^{\mathrm{1s}_{1/2}}_{n, \mathrm{cluster}}$ is blocked, which is marked with crosses.}
\end{figure}

In Fig.~\ref{Fig:Potential} we present the schematic illustration of the single neutron energy levels and their mean-field potential for a cluster $^4$He emersed in a neutron gas with $n_{n,\mathrm{gas}} = 10^{-3}\ \mathrm{fm}^{-3}$, which is fixed adopting the relativistic density functional DD-LZ1~\cite{Wei2020_CPC44-074107}. The orbital 1s$_{1/2}$ is occupied by two neutrons indicated by the solid circles, which form the neutron part of the cluster $^4$He. Due to the effects of Pauli blocking, nucleons in nuclear medium can no longer occupy the orbital 1s$_{1/2}$. The neutron gas indicated by the shaded region remains in constant density with $n_{n,\mathrm{gas}} = 10^{-3}\ \mathrm{fm}^{-3}$, while the local chemical potential $\mu_{n, \mathrm{gas}}(r)$ at small $r$ decreases due to the attractive force exerted by the cluster, i.e.,
\begin{equation}
\mu_{i, \mathrm{gas}}(r) = \sqrt{\nu_i^2+{m_i^*}^2} + \Sigma^\mathrm{R} + g_{\omega} \omega + g_{\rho}\tau_{i} \rho + q_i  A.
\end{equation}
Consequently, the energy gap between $\mu_{n, \mathrm{gas}}(r)$ and the lowest energy state $\varepsilon_{n, \mathrm{gas}}^\mathrm{b}$ in neutron gas decreases in the presence of the cluster. The momentum $\bar{\nu}_i(r)$ corresponding to the lowest gas state $\varepsilon_{n, \mathrm{gas}}^\mathrm{b}$ can be fixed by Eq.~(\ref{eq:nub_gas}), which becomes nonzero with $\bar{\nu}_i(r)>0$. In between the lowest gas state $\varepsilon_{n, \mathrm{gas}}^\mathrm{b}$ and the occupied cluster state $\varepsilon^{\mathrm{1s}_{1/2}}_{n, \mathrm{cluster}}$, there exist another orbital 1p$_{3/2}$, which is blocked (marked with crosses) so that the cluster is stable against absorbing additional nucleons. In principle, for static clusters (e.g., forming crystal structures), they will absorb nucleons and alter the density profiles of nuclear medium with $n_{i,\mathrm{gas}}(r) \neq$ constant so that the system becomes more stable~\cite{Maruyama2005_PRC72-015802}.

The nucleon numbers of clusters can be calculated by integrating the nucleon density $n_{i,\mathrm{cluster}}(r)$ of clusters in coordinate space, i.e.,
\begin{eqnarray}
N_i =   \int 4\pi r^2 n_{i,\mathrm{cluster}}(r) \mbox{d}r,
\end{eqnarray}
where the total mass number is fixed with $A=N_p+N_n$. Based on the proton density profiles, the charge radius $R_\mathrm{ch}$ can be obtained with~\cite{Sugahara1994_NPA579-557}
\begin{equation}
  R_\mathrm{ch}^2 = R_p^2  + (0.862\ \mathrm{fm})^2 - (0.336\ \mathrm{fm})^2\frac{N}{Z}, \label{eq:Rch}
\end{equation}
with the proton radius $R_p$ fixed by
\begin{equation}
  R_p^2 = \frac{1}{Z}\int 4\pi r^4 n_{p,\mathrm{cluster}}(r) \mbox{d}r.
\end{equation}

The energy of the system is obtained with
\begin{equation}
E_\mathrm{MF}=\int \langle {\cal{T}}_{00} \rangle \mbox{d}^3 r, \label{eq:energy}
\end{equation}
where the energy momentum tensor is determined by
\begin{eqnarray}
\langle {\cal{T}}_{00} \rangle
&=&  \sum_{i=n,p}\left[\sum_{k=1}^{N_i} \varepsilon_{ki} \langle \bar{\psi}_{ki} \gamma^{0} \psi_{ki} \rangle+ (m_i -   V_i) n_{i,\mathrm{cluster}}  \right]  \nonumber \\
&&   +\sum_{i=n,p}\mathcal{E}_{i,\mathrm{gas}} + \frac{1}{2}(\nabla \sigma)^2 + \frac{1}{2}m_\sigma^2 \sigma^2 + \frac{1}{2}(\nabla \omega)^2   \nonumber \\
&&   + \frac{1}{2}m_\omega^2 \omega^2 + \frac{1}{2}(\nabla \rho)^2 + \frac{1}{2}m_\rho^2 \rho^2
     + \frac{1}{2}(\nabla A)^2 \nonumber \\
&&   + c_3\omega^4 - U(\sigma, \omega), \label{eq:ener_dens} \\
 \mathcal{E}_{i,\mathrm{gas}} &=& \frac {{m^*_i}^4}{8\pi^{2}}\left[g(x_i) - g(\bar{x}_i) \right],
\end{eqnarray}
with $g(x) = x(2x^2+1)\sqrt{x^2+1}-\mathrm{arcsh}(x)$. The Coulomb energy $E_\mathrm{C}$ can be estimated with
\begin{equation}
E_\mathrm{C} =  \frac{1}{2}\int (\nabla A)^2\mbox{d}^3 r. \label{eq:eC}
\end{equation}
The microscopic center-of-mass correction~\cite{Bender2000_EPJA7-467} is fixed by
\begin{equation}
E_\mathrm{c.m.} = -\frac{\langle P_\mathrm{c.m.}^2 \rangle}{2 \sum_{i=n,p}m_i  N_i}.  \label{eq:ecm}
\end{equation}
The total mass of a cluster emersed in nuclear medium with given $N_i$ and $n_{i,\mathrm{gas}}$ is then fixed by
\begin{equation}
E_\mathrm{tot} = E_\mathrm{MF} + E_\mathrm{c.m.} - E_\mathrm{gas},  \label{eq:etot}
\end{equation}
where $E_\mathrm{gas}$ represents the energy contribution of nuclear medium in the absence of clusters, i.e., the uniform nuclear matter with densities $n_{i,\mathrm{gas}}$.

Finally, in the framework of gRDF~\cite{Typel2010_PRC81-015803}, the binding energy of the cluster is fixed by
\begin{equation}
B = E_\mathrm{tot} - \sum_{i=n,p}(m_i + S_{i,\mathrm{gas}} + V_{i,\mathrm{gas}}) N_i, \label{eq:bind}
\end{equation}
where $S_{i,\mathrm{gas}}$ and $V_{i,\mathrm{gas}}$ are the scalar and vector mean-field potentials of nucleons in the uniform nuclear medium.

\section{\label{sec:res}Results and discussions}
For clusters with fixed neutron $(N_n)$ and proton $(N_p)$ numbers emersed in nuclear medium with density $n_{i,\mathrm{gas}}$, we solve the Dirac Eq.~(\ref{EQ:RDirac}) with mean-field potentials Eqs.~(\ref{eq:MF_scalar}-\ref{eq:re_B}), boson field Eqs.~(\ref{eq:KG_sigma}-\ref{eq:KG_photon}), and densities Eq.~(\ref{eq:np_cluster}-\ref{eq:nub_gas}) in the RMF model iteratively in coordinate space with a box size of $R_\mathrm{W} = 12.8~{\rm fm}$ and a grid distance of $0.1~{\rm fm}$. Once convergency is reached, the energy and binding energy can be fixed by Eqs.~(\ref{eq:etot}) and (\ref{eq:bind}).

\begin{figure}[!ht]
  \centering
  \includegraphics[width=\linewidth]{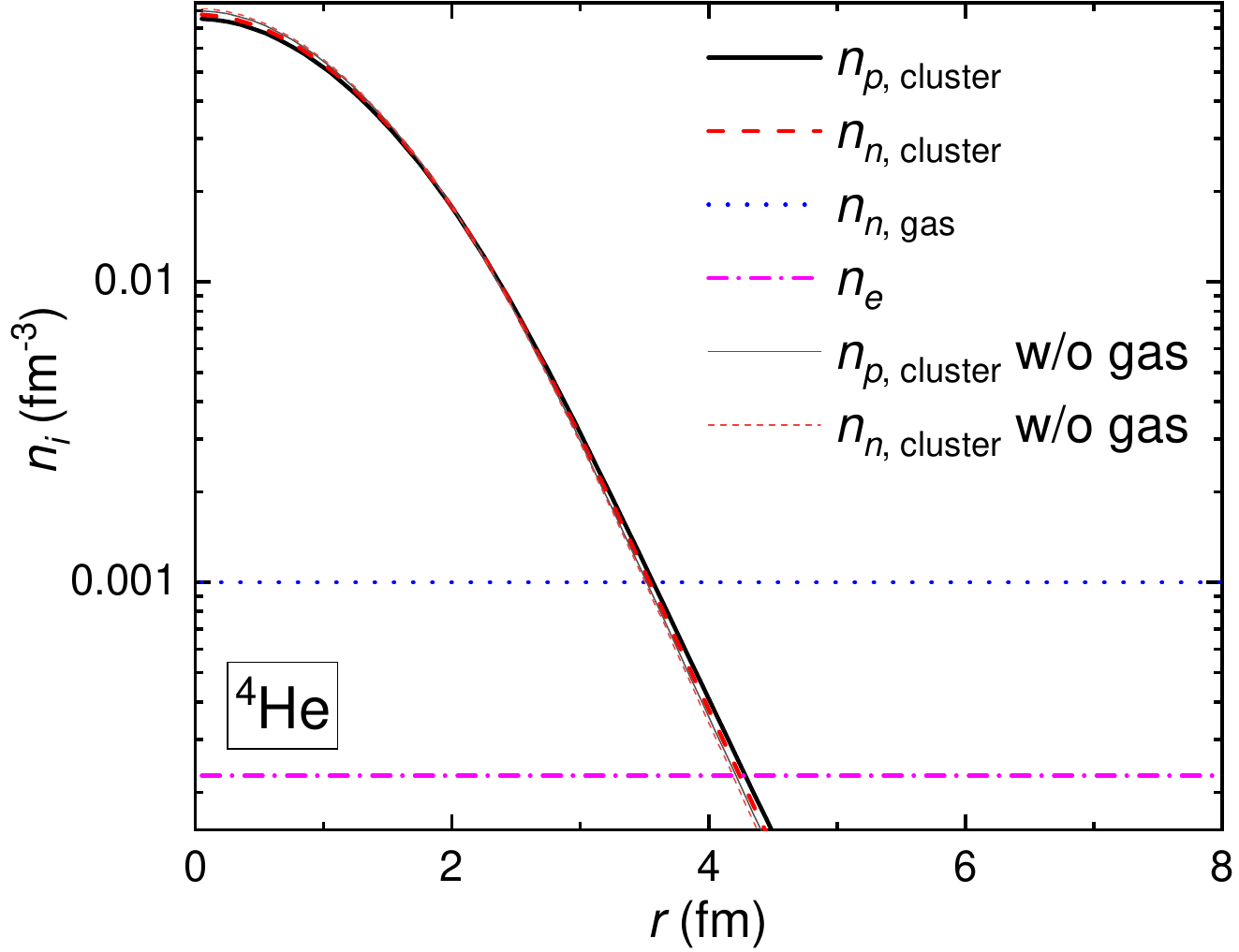}
  \caption{\label{Fig:Dense}Density profiles of $^4$He emersed in a neutron gas with $n_{n,\mathrm{gas}} = 10^{-3}\ \mathrm{fm}^{-3}$, in correspondence to Fig.~\ref{Fig:Potential}. The uniform density $n_e$ is imposed to ensure global charge neutrality for the Wigner-Seitz cell with its size $R_\mathrm{W} = 12.8~{\rm fm}$. The thin curves indicate the density profiles of $^4$He in vacuum.}
\end{figure}

As an example, in Fig.~\ref{Fig:Dense} we present the obtained density profiles of $^{4}$He emersed in a neutron gas with $n_{n,\mathrm{gas}} = 10^{-3}\ \mathrm{fm}^{-3}$, where the corresponding single neutron energy levels and mean-field potential are indicated in Fig.~\ref{Fig:Potential}. A constant electron density $n_{e} = 2.28 \times 10^{-4}\ \mathrm{fm}^{-3}$ is introduced so that the Wigner-Seitz cell with $R_\mathrm{W} = 12.8~{\rm fm}$ carries no charge, which will increase the binding energy due to the reduction of Coulomb energy as indicated in Table~\ref{table:Cluster}.  The density profiles of $^4$He in vacuum are presented and indicated by the thin curves in Fig.~\ref{Fig:Dense}. Evidently, due to the impact of nuclear medium, $^4$He becomes more dilute with its central density reduced, where the binding energy will be altered as well.

\begin{figure}[!ht]
  \centering
  \includegraphics[width=\linewidth]{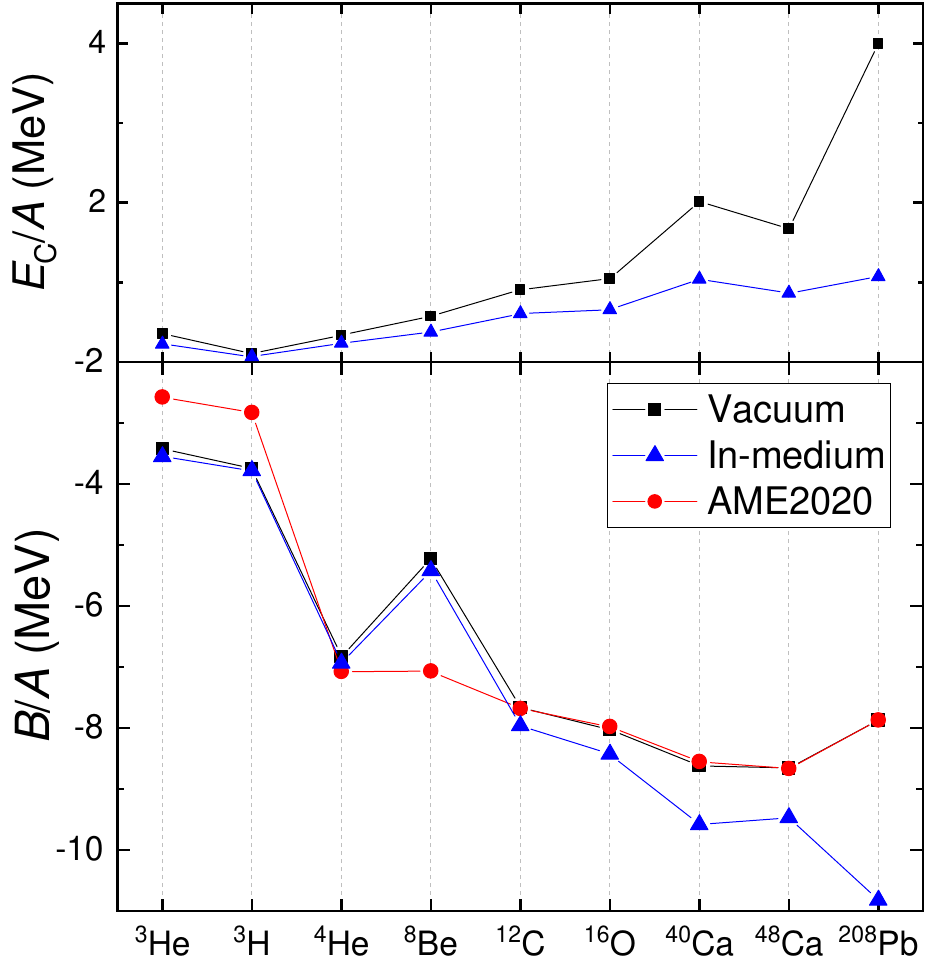}
  \caption{\label{Fig:Bind_comp_DDLZ1}Binding energy $B$ and Coulomb energy $E_\mathrm{C}$ per nucleon for finite nuclei in vacuum and nuclear medium ($n_{n,\mathrm{gas}} = 10^{-10}\ \mathrm{fm}^{-3}$) predicted by the relativistic density functional DD-LZ1~\cite{Wei2020_CPC44-074107}, which are compared with the experimental data from AME2020~\cite{Huang2021_CPC45-30002, Wang2021_CPC45-030003}.}
\end{figure}

To show the effectiveness of RMF models describing both light and heavy nuclei, in Fig.~\ref{Fig:Bind_comp_DDLZ1} we present the binding energy and Coulomb energy per nucleon for finite nuclei in vacuum and nuclear medium, where the relativistic density functional DD-LZ1~\cite{Wei2020_CPC44-074107} was adopted. Evidently, the functional DD-LZ1 well describes the binding energies for nuclei ranging from $^4$He to $^{208}$Pb, while a deviation of $\sim 1$ MeV on $B/A$ is identified for $^3$He and $^3$H. Larger deviation is observed for $^8$Be since we have assumed it to be spherical while in fact it is deformed with $\beta_{20}=1.307$~\cite{Rong2023_PRC108-054314}. If those nuclei are emersed in nuclear medium, as indicated by the blue triangles, large deviations from experimental values are observed with the binding energy increased, where the increment grows and becomes sizable for heavy nuclei. Note that a rather small density $n_{n,\mathrm{gas}} = 10^{-10}\ \mathrm{fm}^{-3}$ for nuclear medium is adopted, where the impact on the properties of clusters is infinitesimal. The increment of binding energy for in-medium clusters is mainly attributed to the background electron gas, where the Coulomb energy $E_\mathrm{C}$ obtained with Eq.~(\ref{eq:eC}) is reduced.

\begin{table}
  \centering
  \caption{\label{table:Cluster} Binding energy $B^\mathrm{th}$, in-medium reduction of Coulomb energy $\Delta E_\mathrm{C}$, and charge radii $R_\mathrm{ch}^\mathrm{th}$ for finite nuclei predicted by the relativistic density functional DD-LZ1~\cite{Wei2020_CPC44-074107}, which are compared with the experimental data $B^\mathrm{exp}$ and $R_\mathrm{ch}^\mathrm{exp}$~\cite{Angeli2013_ADNDT99-69, Huang2021_CPC45-30002, Wang2021_CPC45-030003}. The in-medium reduction of Coulomb energy $\Delta E_\mathrm{C}$ arises due to the inclusion of uniform density $n_e$ to ensure global charge neutrality for the Wigner-Seitz cell with its size $R_\mathrm{W} = 12.8~{\rm fm}$.}
  \begin{tabular}{c|c c|c|cc}
    \hline \hline
  Cluster    &  $B^\mathrm{th}/A$ & $B^\mathrm{exp}/A$& $\Delta E_\mathrm{C}/A$   & $R_\mathrm{ch}^\mathrm{th}$  &  $R_\mathrm{ch}^\mathrm{exp}$\\
             &     MeV     &     MeV   &  MeV  &  fm   &   fm  \\\hline
  $^3$H      &  $-3.747$   & $-2.827$  & 0.033 &  2.374   &   1.759  \\
  $^3$He     &  $-3.419$   & $-2.573$  & 0.131 &  2.864   &   1.966  \\
  $^4$He     &  $-6.832$   & $-7.074$  & 0.100 &  2.252   &   1.675  \\
  $^8$Be     &  $-5.220$   & $-7.062$  & 0.198 &  2.619   &    --    \\
  $^{12}$C   &  $-7.663$   & $-7.680$  & 0.298 &  2.448   &   2.470  \\
  $^{16}$O   &  $-8.028$   & $-7.976$  & 0.395 &  2.752   &   2.699  \\
  $^{40}$Ca  &  $-8.610$   & $-8.551$  & 0.971 &  3.482   &   3.478  \\
  $^{48}$Ca  &  $-8.659$   & $-8.667$  & 0.810 &  3.472   &   3.477  \\
  $^{208}$Pb &  $-7.871$   & $-7.867$  & 2.920 &  5.497   &   5.501  \\   \hline
  \end{tabular}
\end{table}

Meanwhile, the charge radii $R_\mathrm{ch}$ determined by Eq.~(\ref{eq:Rch}) are barely altered by the reduction of Coulomb potential. The obtained binding energy $B^\mathrm{th}$, in-medium reduction of Coulomb energy $\Delta E_\mathrm{C}$, and charge radii $R_\mathrm{ch}^\mathrm{th}$ for finite nuclei are then summarized in Table~\ref{table:Cluster}, where the experimental values for the binding energy $B^\mathrm{exp}$ and charge radii $R_\mathrm{ch}^\mathrm{exp}$ are presented as well~\cite{Angeli2013_ADNDT99-69, Huang2021_CPC45-30002, Wang2021_CPC45-030003}. Note that the obtained $R_\mathrm{ch}$ is slightly larger than previous estimations since we have adopted reflective boundary conditions for boson fields and Dirichlet-Neumann boundary condition for nucleons~\cite{Negele1973_NPA207-298} with $R_\mathrm{W}=12.8$ fm.

\begin{figure}[!ht]
  \centering
  \includegraphics[width=0.8\linewidth]{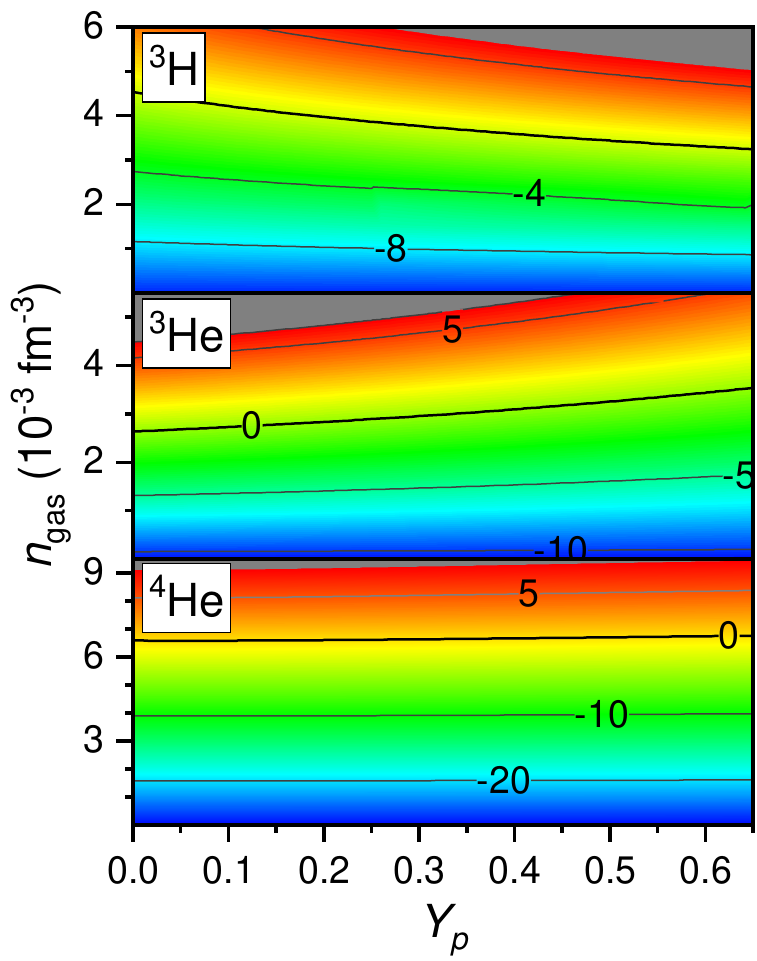}
  \caption{\label{Fig:DBall_DDLZ1} Binding energies of light clusters emersed in nuclear medium with density $n_{\mathrm{gas}}=n_{p,\mathrm{gas}}+n_{n,\mathrm{gas}}$ and proton fraction $Y_p = n_{p,\mathrm{gas}}/n_{\mathrm{gas}}$, which are obtained by employing the relativistic density functional DD-LZ1~\cite{Wei2020_CPC44-074107}.}
\end{figure}

By increasing the densities of nuclear medium, we can then examine the variations on the properties of various clusters in nuclear medium. In Fig.~\ref{Fig:DBall_DDLZ1} we present the binding energy per nucleon for light clusters emersed in nuclear medium with density $n_{\mathrm{gas}}=n_{p,\mathrm{gas}}+n_{n,\mathrm{gas}}$ and proton fraction $Y_p = n_{p,\mathrm{gas}}/n_{\mathrm{gas}}$, which are obtained by employing the relativistic density functional DD-LZ1~\cite{Wei2020_CPC44-074107}. At vanishing densities for nuclear medium, as indicated in Table~\ref{table:Cluster}, in-medium nuclei is more bound than those in vacuum due to the reduction of Coulomb energy. As $n_{\mathrm{gas}}$ increases, the binding energies of clusters decrease, which would eventually become unstable and melt at sufficient large densities, i.e., Mott transition. The variation of binding energies with respect to $Y_p$ can also be identified, which are sensitive to the differences between neutron and proton numbers. In particular, the mirror nuclear pair $^3$H and $^3$He shows opposite trends, where the binding energy of $^3$He increases with $Y_p$ and that of $^3$H decreases. This may be useful to estimate the symmetry energy of nuclear matter by examine the respective yields in heavy-ion collisions~\cite{Custodio2025_PRL134-082304}. Meanwhile, the binding energy of $^4$He varies little with respect to $Y_p$ since $N_p=N_n=2$.

\begin{table}
  \centering
  \caption{\label{table:bind_exp} Coefficients for the expansion formula (\ref{eq:bind_exp}) of binding energy for various clusters presented in Fig.~\ref{Fig:DBall_DDLZ1} and Fig.~\ref{Fig:Bind-DDLZ1}, where the relativistic density functional DD-LZ1 is employed~\cite{Wei2020_CPC44-074107}.}
  \begin{tabular}{c|cccccc}
    \hline \hline
               &  $B_0$       & $a$       & $b$   &   $c$          &  $d$                 &  $f$    \\
             &  MeV         & GeV$\cdot$fm$^3$ &  GeV$\cdot$fm$^6$ &  GeV$\cdot$fm$^3$ & MeV &   MeV   \\\hline
  $^3$H      &  $-11.53$    & 3.089 &  $-113.3$  &  1.113      & 1.17     & $-1.06$  \\
  $^3$He     &  $-10.57$    & 4.388 &  $-149.2$  &  $-1.197$   & 0.26     & $-1.02$   \\
  $^4$He     &  $-27.702$   & 5.029 &  $-122.3$  &  $-0.166$   & 0.60     & $-0.77$  \\ \hline
  $^8$Be     &  $-43.34$    & 10.06 &   $-262$   &  $-0.268$   & $-0.10$  &   0     \\
  $^{12}$C   &  $-95.40$    & 14.55 &   $-221$   &  $-1.03$    &  0.37    &   0     \\
  $^{16}$O   & $-134.62$    & 19.32 &   $-320$   &  $-1.41$    &  0.44    &   0     \\
  $^{40}$Ca  &  $-383.03$   & 48.47 &   $-714$   &  $-4.53$    &   1.3    &   0     \\
  $^{48}$Ca  &  $-454.2$    & 52.32 &   $-712$   &    4.8      &   3.4    &   0     \\
  $^{208}$Pb &  $-2248.5$   & 220.6 &   $-2581$  &    28.2     &   23     &   0     \\ \hline
  \end{tabular}
\end{table}

The variation of binding energy with respect to the density $n_{\mathrm{gas}}$ and proton fraction $Y_p$ of nuclear medium can be fitted with the following polynomial formula, i.e.,
\begin{equation}
  B = B_0  + a n_{\mathrm{gas}}  + b n_{\mathrm{gas}}^2 + c Y_p n_{\mathrm{gas}} + d Y_p + f Y_p^2, \label{eq:bind_exp}
\end{equation}
where the corresponding coefficients are indicated in Table~\ref{table:bind_exp}. The coefficients $c$, $d$, $f$ give the dependence of binding energy on the isospin asymmetry of nuclear medium. Particularly, corresponding to different trends of binding energy with respect to $Y_p$, $c$ takes opposite signs for $^3$H and $^3$He, while that of $^4$He takes a relatively small value. Note that the binding energies at $n_{\mathrm{gas}}=Y_p=0$ generally fulfill the relation $B_0 = B^\mathrm{th} - \Delta E_\mathrm{C}$ with $B^\mathrm{th}$ and $\Delta E_\mathrm{C}$ listed in Table~\ref{table:Cluster}.

\begin{figure}[!ht]
  \centering
  \includegraphics[width=0.85\linewidth]{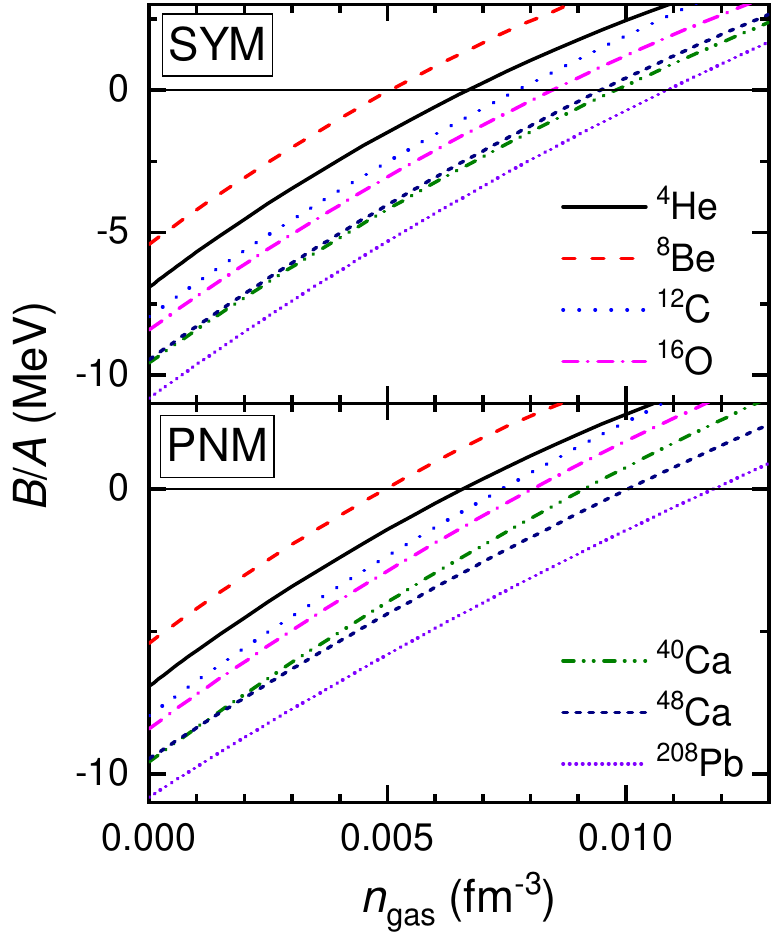}
  \caption{\label{Fig:Bind-DDLZ1} Binding energy per nucleon for various clusters emersed in symmetric nuclear matter (SYM) and pure neutron matter (PNM) as functions of the density of nuclear medium $n_{\mathrm{gas}}=n_{p,\mathrm{gas}}+n_{n,\mathrm{gas}}$, which are obtained by employing the relativistic density functional DD-LZ1~\cite{Wei2020_CPC44-074107}.}
\end{figure}

In Fig.~\ref{Fig:Bind-DDLZ1} we present the binding energy per nucleon for heavy clusters emersed in symmetric nuclear matter (SYM, $Y_p = 0.5$) and pure neutron matter (PNM, $Y_p = 0$) as functions of the density of nuclear medium $n_{\mathrm{gas}}$. Similar to light clusters, the binding energies of all clusters decrease with $n_{\mathrm{gas}}$. The variations in binding energies with respect to $Y_p$ can be identified by comparing the upper and lower panels in Fig.~\ref{Fig:Bind-DDLZ1}. Similar to light clusters indicated in Fig.~\ref{Fig:DBall_DDLZ1}, nuclei with $N_p=N_n$ are generally insensitive to the variations in $Y_p$, while that of $^{48}$Ca and $^{208}$Pb are altered. Based on the results indicated in Fig.~\ref{Fig:Bind-DDLZ1}, we can also fit the binding energies using Eq.~(\ref{eq:bind_exp}), where the coefficients are indicated in Table~\ref{table:bind_exp} as well. Note that we have omitted the $Y_p^2$ term by taking $f=0$. Finally, as stressed earlier, the quantum fluctuations of heavy clusters will eventually become insignificant due to their large masses, then the formation of crystal structures becomes favorable at vanishing temperatures. In such cases, the assumption of a uniform background density for nuclear medium is inappropriate and the number of nucleons in clusters will be altered as well~\cite{Maruyama2005_PRC72-015802}.

\begin{figure}[!ht]
  \centering
  \includegraphics[width=0.8\linewidth]{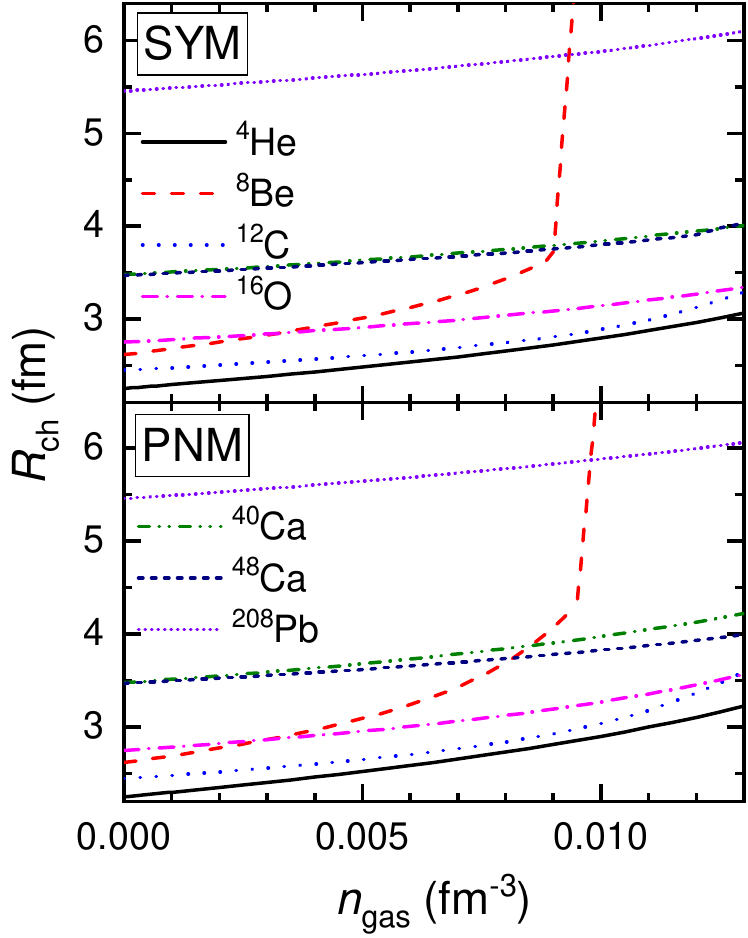}
  \caption{\label{Fig:Rch-DDLZ1} Similar as Fig.~\ref{Fig:Bind-DDLZ1} but for charge radii $R_\mathrm{ch}$ of in-medium clusters.}
\end{figure}

In Fig.~\ref{Fig:Rch-DDLZ1} we present the charge radii $R_\mathrm{ch}$ of in-medium clusters predicted by the relativistic density functional DD-LZ1~\cite{Wei2020_CPC44-074107}. In general, we find $R_\mathrm{ch}$ increases with the density of nuclear medium $n_{\mathrm{gas}}$. Particularly, the charge radii of $^8$Be grow significantly at $n_{\mathrm{gas}}\approx 0.0095\ \mathrm{fm}^{-3}$, where the highest occupied single nucleon level $\varepsilon^{\mathrm{1p}_{3/2}}_{n, \mathrm{cluster}}$ of the cluster becomes a continuum state and overlaps with the lowest gas state $\varepsilon_{n, \mathrm{gas}}^\mathrm{b}$, i.e., the orbital 1p$_{3/2}$ becomes unbound and extends throughout the Wigner-Seitz cell. The cluster $^8$Be then becomes unbound and melts in nuclear medium. The variations of $R_\mathrm{ch}$ with respect to $Y_p$ can be identified by comparing the upper and lower panels in Fig.~\ref{Fig:Rch-DDLZ1}, where $R_\mathrm{ch}$ increases slightly faster for clusters with $N_p=N_n$ in PNM than those in SYM as we increase $n_{\mathrm{gas}}$. This could be attributed to the attractive interaction between the protons in clusters and the neutron gas. For clusters with $N_p<N_n$ such as $^{48}$Ca and $^{208}$Pb, the variations of $R_\mathrm{ch}$ with respect to $Y_p$ are insignificant. It is found that the charge radii for $^{48}$Ca and $^{208}$Pb are slightly larger in SYM than those in PNM, which is the opposite for clusters with $N_p=N_n$.

\begin{figure*}
\includegraphics[width=0.38\linewidth]{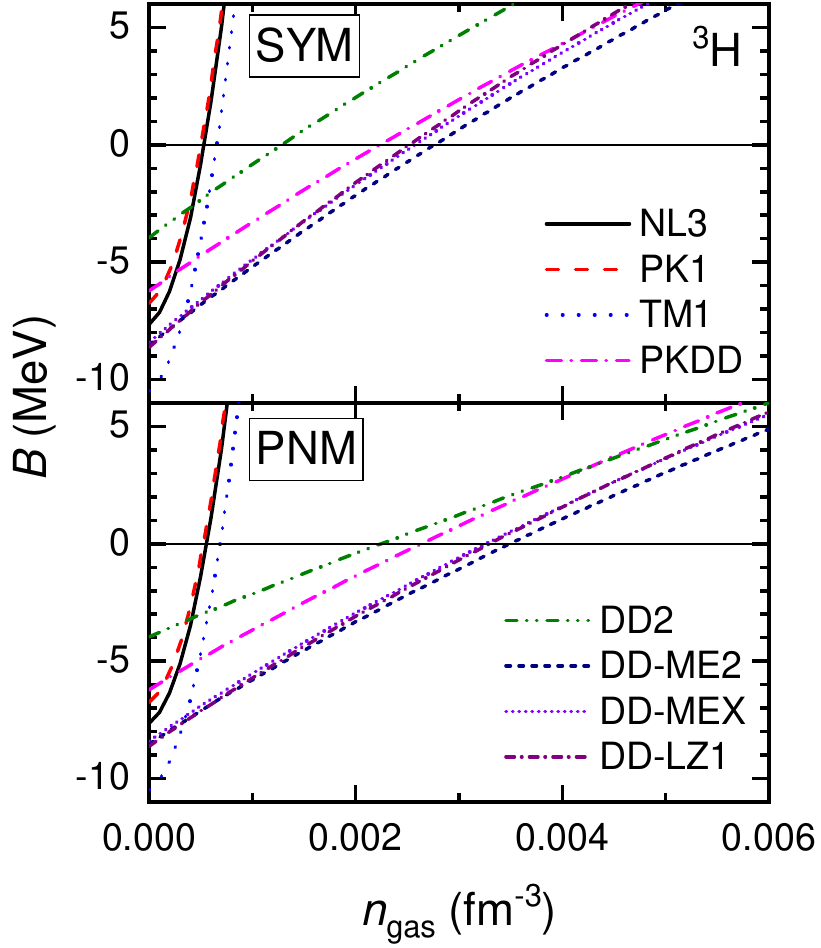} \ \ \  \ \ \
\includegraphics[width=0.36\linewidth]{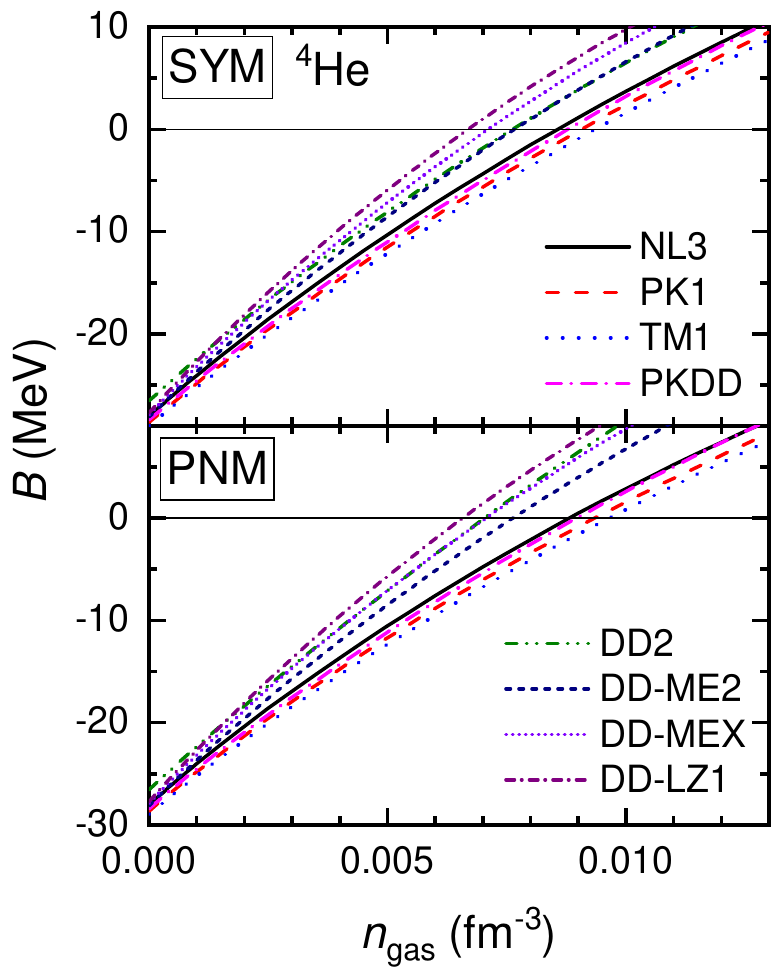}
  \caption{\label{Fig:Bind_HeH} Binding energies of $^3$H and $^4$He emersed in SYM and PNM as functions of $n_{\mathrm{gas}}$, which are obtained adopting various relativistic density functionals. The results are fitted with Eq.~(\ref{eq:bind_exp}) with the coefficients indicated in Table.~\ref{table:bind_exp_L}.   }
\end{figure*}

Finally, to show the impact of uncertainties in nuclear matter properties on the binding energies of in-medium clusters, we examine the binding energies of $^3$H and $^4$He emersed in SYM and PNM adopting various relativistic density functionals, i.e., NL3~\cite{Lalazissis1997_PRC55-540}, PK1~\cite{Long2004_PRC69-034319}, TM1~\cite{Sugahara1994_NPA579-557}, PKDD~\cite{Long2004_PRC69-034319}, DD2~\cite{Typel2010_PRC81-015803}, DD-ME2~\cite{Lalazissis2005_PRC71-024312}, DD-MEX~\cite{Taninah2020_PLB800-135065}, and DD-LZ1~\cite{Wei2020_CPC44-074107}. The obtained results are then presented in in Fig.~\ref{Fig:Bind_HeH}. At vanishing densities, the binding energy of $^4$He is well reproduced by all functionals, while there exist deviations ($\sim$3 MeV) for the binding energy of $^3$H except for DD-ME2, DD-MEX, and DD-LZ1. The binding energies for $^3$H and $^4$He are generally increasing with the density of nuclear medium $n_{\mathrm{gas}}$. As observed earlier in Fig.~\ref{Fig:DBall_DDLZ1}, the binding energies of $^3$H decrease with the proton fraction $Y_p$ of nuclear medium, while that of $^4$He is insensitive to the variations of $Y_p$. The uncertainties on the binding energies of in-medium $^3$H are much larger than that of $^4$He. In particular, the binding energies of $^3$H decrease drastically at $n_{\mathrm{gas}}\approx 0.0008\ \mathrm{fm}^{-3}$ when we adopt non-linear relativistic density functionals NL3~\cite{Lalazissis1997_PRC55-540}, PK1~\cite{Long2004_PRC69-034319}, and TM1~\cite{Sugahara1994_NPA579-557}, while the density-dependent functionals PKDD~\cite{Long2004_PRC69-034319}, DD2~\cite{Typel2010_PRC81-015803}, DD-ME2~\cite{Lalazissis2005_PRC71-024312}, DD-MEX~\cite{Taninah2020_PLB800-135065}, and DD-LZ1~\cite{Wei2020_CPC44-074107} predict relatively slow decline for the binding energies.

\begin{table}
  \centering
  \caption{\label{table:bind_exp_L} Coefficients for the expansion formula (\ref{eq:bind_exp}) of binding energies for
  $^3$H and $^4$He presented in Fig.~\ref{Fig:Bind_HeH}, where the relativistic density functionals NL3~\cite{Lalazissis1997_PRC55-540}, PK1~\cite{Long2004_PRC69-034319}, TM1~\cite{Sugahara1994_NPA579-557}, PKDD~\cite{Long2004_PRC69-034319}, DD2~\cite{Typel2010_PRC81-015803}, DD-ME2~\cite{Lalazissis2005_PRC71-024312}, and DD-MEX~\cite{Taninah2020_PLB800-135065} have been adopted. Here we take $f=0$.}
  \begin{tabular}{c|c|ccccc}
    \hline \hline
  &           &  $B_0$       & $a$       & $b$   &   $c$          &  $d$                     \\
  &           &  MeV         & GeV$\cdot$fm$^3$ &  GeV$\cdot$fm$^6$ &  GeV$\cdot$fm$^3$ & MeV    \\ \hline
\multirow{7}*{$^3$H} &  NL3      &  $-7.632$    & 2.05  &  20886   &   2.49     & $-0.08$      \\
  &  PK1      &  $-6.746$    & 1.86  &  20752   &   2.48     & $-0.08$      \\
  &  TM1      &  $-10.47$    & 1.70  &  19761   &   30     & $-1.4$      \\
  &  PKDD     &  $-6.20$     & 2.56  &  $-77$   &   0.770  & $-0.002$      \\
  & DD2       &  $-3.939$    & 1.795 &  $-22$   &   2.14     & 0.20        \\
  & DD-ME2    &  $-8.49$     & 2.71  &  $-79$   &   1.06     & 0.08        \\
  & DD-MEX    &  $-8.37$     & 2.87  &   $-91$  &   1.11     & 0.08        \\ \hline
\multirow{7}*{$^4$He} &  NL3      &  $-27.81$    & 3.71  &  $-63.6$   &   0.223     & $-0.32$      \\
 &  PK1      &  $-28.46$    & 3.58  &  $-57.0$   &   0.213     & $-0.30$      \\
 &  TM1      &  $-28.78$    & 3.51  &  $-53.8$   &   0.205     & $-0.31$      \\
 &  PKDD     &  $-28.24$    & 3.65  &  $-55.6$   &   0.144     & $-0.22$      \\
 & DD2       &  $-26.62$    & 4.26 &   $-64.5$   &  $-0.691$    & 0.63        \\
 & DD-ME2    &  $-27.98$    & 4.26 &   $-77.3$   &  $-0.058$    & 0.04        \\
 & DD-MEX    &  $-27.88$    & 4.68 &   $-100$    &  $-0.067$    & 0.04        \\ \hline
 \end{tabular}
\end{table}

Based on the variations of binding energies indicated in Fig.~\ref{Fig:Bind_HeH}, similar to Table~\ref{table:bind_exp}, we fit the obtained results using Eq.~(\ref{eq:bind_exp}) and present the coefficients in Table~\ref{table:bind_exp_L}. The binding energies $B_0$ for $^3$H in vanishing densities varies with the adopted density functionals, while those of $^4$He are generally in consistent with the experimental data indicated in Table~\ref{table:Cluster}. Slight variations on the parameters $a$ and $b$ describing the density dependence of binding energies are observed for $^4$He, which become significant for $^3$H. For the parameters $c$ and $d$ describing the dependence on $Y_p$, they are relatively small for $^4$He except for DD2, while $c$ and $d$ are altered for $^3$H when we adopt different functionals.

\section{\label{sec:con}Conclusion}

In this work, we propose a hybrid treatment for investigating the clustering phenomenon in nuclear medium in the framework of relativistic-mean-field models. Assuming a spherically symmetric Wigner-Seitz cell, the clusters are fixed by solving the Dirac equations imposing Dirichlet-Neumann boundary condition, while a reflective boundary condition is adopted for boson fields. The nuclear medium are treated with Thomas-Fermi approximation and take constant densities. By blocking the energy states in between the lowest gas sate and the last occupied cluster state, the cluster becomes stable against absorbing additional nucleons in nuclear medium.

We then investigate the properties of both light and heavy clusters emersed in nuclear medium at various densities $n_{\mathrm{gas}}$ and proton fractions $Y_p$. Our findings are summarized as follows.
\begin{itemize}
  \item The RMF models generally well reproduce the binding energies for $^4$He, $^{12}$C, $^{16}$O, $^{40}$Ca, $^{48}$Ca, and $^{208}$Pb, while slight deviations are observed for $^3$H, $^3$He, and $^8$Be adopting certain relativistic density functionals.
  \item By introducing a background electron density, the binding energy of clusters increases due to the reduction of Coulomb energy.
  \item As the density of nuclear medium increases, the clusters eventually become unbound with the root-mean-square charge radii increased.
  \item The binding energies of in-medium clusters increase (decrease) with the proton fraction $Y_p$ of nuclear medium if they have more (less) protons than neutrons, while those with same proton and neutron numbers $N_p = N_n$ are generally insensitive to $Y_p$.
  \item The uncertainties in nuclear matter properties have more significant impact on the binding energies of in-medium clusters with $N_p \neq N_n$ than those with $N_p = N_n$.
  \item The variation of binding energies with respect to the density and proton fraction of nuclear medium are fitted with Eq.~(\ref{eq:bind_exp}), where the coefficients are presented in Table~\ref{table:bind_exp} and \ref{table:bind_exp_L}.
\end{itemize}
The results presented in this work should be useful to unveil the clustering phenomenon in both heavy-ion collisions and neutron stars.

\begin{acknowledgments}
The author would like to thank Prof. Bing-Nan Lu, Prof. Yu-Ting Rong, and Prof. Ting-Ting Sun for fruitful discussions.
This work was supported by the National Natural Science Foundation of China (Grant No. 12275234) and the National SKA Program of China (Grant No. 2020SKA0120300).
\end{acknowledgments}


\begin{thebibliography}{67}%
\makeatletter
\providecommand \@ifxundefined [1]{%
 \@ifx{#1\undefined}
}%
\providecommand \@ifnum [1]{%
 \ifnum #1\expandafter \@firstoftwo
 \else \expandafter \@secondoftwo
 \fi
}%
\providecommand \@ifx [1]{%
 \ifx #1\expandafter \@firstoftwo
 \else \expandafter \@secondoftwo
 \fi
}%
\providecommand \natexlab [1]{#1}%
\providecommand \enquote  [1]{``#1''}%
\providecommand \bibnamefont  [1]{#1}%
\providecommand \bibfnamefont [1]{#1}%
\providecommand \citenamefont [1]{#1}%
\providecommand \href@noop [0]{\@secondoftwo}%
\providecommand \href [0]{\begingroup \@sanitize@url \@href}%
\providecommand \@href[1]{\@@startlink{#1}\@@href}%
\providecommand \@@href[1]{\endgroup#1\@@endlink}%
\providecommand \@sanitize@url [0]{\catcode `\\12\catcode `\$12\catcode
  `\&12\catcode `\#12\catcode `\^12\catcode `\_12\catcode `\%12\relax}%
\providecommand \@@startlink[1]{}%
\providecommand \@@endlink[0]{}%
\providecommand \url  [0]{\begingroup\@sanitize@url \@url }%
\providecommand \@url [1]{\endgroup\@href {#1}{\urlprefix }}%
\providecommand \urlprefix  [0]{URL }%
\providecommand \Eprint [0]{\href }%
\providecommand \doibase [0]{http://dx.doi.org/}%
\providecommand \selectlanguage [0]{\@gobble}%
\providecommand \bibinfo  [0]{\@secondoftwo}%
\providecommand \bibfield  [0]{\@secondoftwo}%
\providecommand \translation [1]{[#1]}%
\providecommand \BibitemOpen [0]{}%
\providecommand \bibitemStop [0]{}%
\providecommand \bibitemNoStop [0]{.\EOS\space}%
\providecommand \EOS [0]{\spacefactor3000\relax}%
\providecommand \BibitemShut  [1]{\csname bibitem#1\endcsname}%
\let\auto@bib@innerbib\@empty
\bibitem [{\citenamefont {R\"opke}(2015)}]{Roepke2015_arXiv1501-01222}%
  \BibitemOpen
  \bibfield  {author} {\bibinfo {author} {\bibfnamefont {G.}~\bibnamefont
  {R\"opke}},\ }\href {https://arxiv.org/abs/1501.01222} {\bibfield  {journal}
  {\bibinfo  {journal} {arXiv:1501.01222}\ } (\bibinfo {year}
  {2015})}\BibitemShut {NoStop}%
\bibitem [{\citenamefont {Lu}\ \emph {et~al.}(2020)\citenamefont {Lu},
  \citenamefont {Li}, \citenamefont {Elhatisari}, \citenamefont {Lee},
  \citenamefont {Drut}, \citenamefont {L\"ahde}, \citenamefont {Epelbaum},\
  and\ \citenamefont {Mei\ss{}ner}}]{Lu2020_PRL125-192502}%
  \BibitemOpen
  \bibfield  {author} {\bibinfo {author} {\bibfnamefont {B.-N.}\ \bibnamefont
  {Lu}}, \bibinfo {author} {\bibfnamefont {N.}~\bibnamefont {Li}}, \bibinfo
  {author} {\bibfnamefont {S.}~\bibnamefont {Elhatisari}}, \bibinfo {author}
  {\bibfnamefont {D.}~\bibnamefont {Lee}}, \bibinfo {author} {\bibfnamefont
  {J.~E.}\ \bibnamefont {Drut}}, \bibinfo {author} {\bibfnamefont {T.~A.}\
  \bibnamefont {L\"ahde}}, \bibinfo {author} {\bibfnamefont {E.}~\bibnamefont
  {Epelbaum}}, \ and\ \bibinfo {author} {\bibfnamefont {U.-G.}\ \bibnamefont
  {Mei\ss{}ner}},\ }\href {\doibase 10.1103/PhysRevLett.125.192502} {\bibfield
  {journal} {\bibinfo  {journal} {Phys. Rev. Lett.}\ }\textbf {\bibinfo
  {volume} {125}},\ \bibinfo {pages} {192502} (\bibinfo {year}
  {2020})}\BibitemShut {NoStop}%
\bibitem [{\citenamefont {Ebran}\ \emph {et~al.}(2012)\citenamefont {Ebran},
  \citenamefont {Khan}, \citenamefont {Niksic},\ and\ \citenamefont
  {Vretenar}}]{Ebran2012_Nature487-341}%
  \BibitemOpen
  \bibfield  {author} {\bibinfo {author} {\bibfnamefont {J.-P.}\ \bibnamefont
  {Ebran}}, \bibinfo {author} {\bibfnamefont {E.}~\bibnamefont {Khan}},
  \bibinfo {author} {\bibfnamefont {T.}~\bibnamefont {Niksic}}, \ and\ \bibinfo
  {author} {\bibfnamefont {D.}~\bibnamefont {Vretenar}},\ }\href {\doibase
  10.1038/nature11246} {\bibfield  {journal} {\bibinfo  {journal} {Nature}\
  }\textbf {\bibinfo {volume} {487}},\ \bibinfo {pages} {341} (\bibinfo {year}
  {2012})}\BibitemShut {NoStop}%
\bibitem [{\citenamefont {R\"opke}\ \emph {et~al.}(2014)\citenamefont
  {R\"opke}, \citenamefont {Schuck}, \citenamefont {Funaki}, \citenamefont
  {Horiuchi}, \citenamefont {Ren}, \citenamefont {Tohsaki}, \citenamefont {Xu},
  \citenamefont {Yamada},\ and\ \citenamefont
  {Zhou}}]{Roepke2014_PRC90-034304}%
  \BibitemOpen
  \bibfield  {author} {\bibinfo {author} {\bibfnamefont {G.}~\bibnamefont
  {R\"opke}}, \bibinfo {author} {\bibfnamefont {P.}~\bibnamefont {Schuck}},
  \bibinfo {author} {\bibfnamefont {Y.}~\bibnamefont {Funaki}}, \bibinfo
  {author} {\bibfnamefont {H.}~\bibnamefont {Horiuchi}}, \bibinfo {author}
  {\bibfnamefont {Z.}~\bibnamefont {Ren}}, \bibinfo {author} {\bibfnamefont
  {A.}~\bibnamefont {Tohsaki}}, \bibinfo {author} {\bibfnamefont
  {C.}~\bibnamefont {Xu}}, \bibinfo {author} {\bibfnamefont {T.}~\bibnamefont
  {Yamada}}, \ and\ \bibinfo {author} {\bibfnamefont {B.}~\bibnamefont
  {Zhou}},\ }\href {\doibase 10.1103/PhysRevC.90.034304} {\bibfield  {journal}
  {\bibinfo  {journal} {Phys. Rev. C}\ }\textbf {\bibinfo {volume} {90}},\
  \bibinfo {pages} {034304} (\bibinfo {year} {2014})}\BibitemShut {NoStop}%
\bibitem [{\citenamefont {Xu}\ \emph {et~al.}(2016)\citenamefont {Xu},
  \citenamefont {Ren}, \citenamefont {R\"opke}, \citenamefont {Schuck},
  \citenamefont {Funaki}, \citenamefont {Horiuchi}, \citenamefont {Tohsaki},
  \citenamefont {Yamada},\ and\ \citenamefont {Zhou}}]{Xu2016_PRC93-011306}%
  \BibitemOpen
  \bibfield  {author} {\bibinfo {author} {\bibfnamefont {C.}~\bibnamefont
  {Xu}}, \bibinfo {author} {\bibfnamefont {Z.}~\bibnamefont {Ren}}, \bibinfo
  {author} {\bibfnamefont {G.}~\bibnamefont {R\"opke}}, \bibinfo {author}
  {\bibfnamefont {P.}~\bibnamefont {Schuck}}, \bibinfo {author} {\bibfnamefont
  {Y.}~\bibnamefont {Funaki}}, \bibinfo {author} {\bibfnamefont
  {H.}~\bibnamefont {Horiuchi}}, \bibinfo {author} {\bibfnamefont
  {A.}~\bibnamefont {Tohsaki}}, \bibinfo {author} {\bibfnamefont
  {T.}~\bibnamefont {Yamada}}, \ and\ \bibinfo {author} {\bibfnamefont
  {B.}~\bibnamefont {Zhou}},\ }\href {\doibase 10.1103/PhysRevC.93.011306}
  {\bibfield  {journal} {\bibinfo  {journal} {Phys. Rev. C}\ }\textbf {\bibinfo
  {volume} {93}},\ \bibinfo {pages} {011306} (\bibinfo {year}
  {2016})}\BibitemShut {NoStop}%
\bibitem [{\citenamefont {Tohsaki}\ \emph {et~al.}(2017)\citenamefont
  {Tohsaki}, \citenamefont {Horiuchi}, \citenamefont {Schuck},\ and\
  \citenamefont {R\"opke}}]{Tohsaki2017_RMP89-011002}%
  \BibitemOpen
  \bibfield  {author} {\bibinfo {author} {\bibfnamefont {A.}~\bibnamefont
  {Tohsaki}}, \bibinfo {author} {\bibfnamefont {H.}~\bibnamefont {Horiuchi}},
  \bibinfo {author} {\bibfnamefont {P.}~\bibnamefont {Schuck}}, \ and\ \bibinfo
  {author} {\bibfnamefont {G.}~\bibnamefont {R\"opke}},\ }\href {\doibase
  10.1103/RevModPhys.89.011002} {\bibfield  {journal} {\bibinfo  {journal}
  {Rev. Mod. Phys.}\ }\textbf {\bibinfo {volume} {89}},\ \bibinfo {pages}
  {011002} (\bibinfo {year} {2017})}\BibitemShut {NoStop}%
\bibitem [{\citenamefont {Xia}\ \emph {et~al.}(2022)\citenamefont {Xia},
  \citenamefont {Maruyama}, \citenamefont {Li}, \citenamefont {Sun},
  \citenamefont {Long},\ and\ \citenamefont {Zhang}}]{Xia2022_CTP74-095303}%
  \BibitemOpen
  \bibfield  {author} {\bibinfo {author} {\bibfnamefont {C.-J.}\ \bibnamefont
  {Xia}}, \bibinfo {author} {\bibfnamefont {T.}~\bibnamefont {Maruyama}},
  \bibinfo {author} {\bibfnamefont {A.}~\bibnamefont {Li}}, \bibinfo {author}
  {\bibfnamefont {B.~Y.}\ \bibnamefont {Sun}}, \bibinfo {author} {\bibfnamefont
  {W.-H.}\ \bibnamefont {Long}}, \ and\ \bibinfo {author} {\bibfnamefont
  {Y.-X.}\ \bibnamefont {Zhang}},\ }\href {\doibase 10.1088/1572-9494/ac71fd}
  {\bibfield  {journal} {\bibinfo  {journal} {Commun. Theor. Phys.}\ }\textbf
  {\bibinfo {volume} {74}},\ \bibinfo {pages} {095303} (\bibinfo {year}
  {2022})}\BibitemShut {NoStop}%
\bibitem [{\citenamefont {Pais}\ \emph {et~al.}(2025)\citenamefont {Pais},
  \citenamefont {Dinh-Thi}, \citenamefont {Fantina}, \citenamefont
  {Gulminelli},\ and\ \citenamefont {Provid\^encia}}]{Pais2025}%
  \BibitemOpen
  \bibfield  {author} {\bibinfo {author} {\bibfnamefont {H.}~\bibnamefont
  {Pais}}, \bibinfo {author} {\bibfnamefont {H.}~\bibnamefont {Dinh-Thi}},
  \bibinfo {author} {\bibfnamefont {A.~F.}\ \bibnamefont {Fantina}}, \bibinfo
  {author} {\bibfnamefont {F.}~\bibnamefont {Gulminelli}}, \ and\ \bibinfo
  {author} {\bibfnamefont {C.}~\bibnamefont {Provid\^encia}},\ }\href@noop {}
  {\  (\bibinfo {year} {2025})},\ \Eprint {http://arxiv.org/abs/2501.12064}
  {arXiv:2501.12064 [nucl-th]} \BibitemShut {NoStop}%
\bibitem [{\citenamefont {Qin}\ \emph {et~al.}(2012)\citenamefont {Qin},
  \citenamefont {Hagel}, \citenamefont {Wada}, \citenamefont {Natowitz},
  \citenamefont {Shlomo}, \citenamefont {Bonasera}, \citenamefont {R\"opke},
  \citenamefont {Typel}, \citenamefont {Chen}, \citenamefont {Huang},
  \citenamefont {Wang}, \citenamefont {Zheng}, \citenamefont {Kowalski},
  \citenamefont {Barbui}, \citenamefont {Rodrigues}, \citenamefont {Schmidt},
  \citenamefont {Fabris}, \citenamefont {Lunardon}, \citenamefont {Moretto},
  \citenamefont {Nebbia}, \citenamefont {Pesente}, \citenamefont {Rizzi},
  \citenamefont {Viesti}, \citenamefont {Cinausero}, \citenamefont {Prete},
  \citenamefont {Keutgen}, \citenamefont {El~Masri}, \citenamefont {Majka},\
  and\ \citenamefont {Ma}}]{Qin2012_PRL108-172701}%
  \BibitemOpen
  \bibfield  {author} {\bibinfo {author} {\bibfnamefont {L.}~\bibnamefont
  {Qin}}, \bibinfo {author} {\bibfnamefont {K.}~\bibnamefont {Hagel}}, \bibinfo
  {author} {\bibfnamefont {R.}~\bibnamefont {Wada}}, \bibinfo {author}
  {\bibfnamefont {J.~B.}\ \bibnamefont {Natowitz}}, \bibinfo {author}
  {\bibfnamefont {S.}~\bibnamefont {Shlomo}}, \bibinfo {author} {\bibfnamefont
  {A.}~\bibnamefont {Bonasera}}, \bibinfo {author} {\bibfnamefont
  {G.}~\bibnamefont {R\"opke}}, \bibinfo {author} {\bibfnamefont
  {S.}~\bibnamefont {Typel}}, \bibinfo {author} {\bibfnamefont
  {Z.}~\bibnamefont {Chen}}, \bibinfo {author} {\bibfnamefont {M.}~\bibnamefont
  {Huang}}, \bibinfo {author} {\bibfnamefont {J.}~\bibnamefont {Wang}},
  \bibinfo {author} {\bibfnamefont {H.}~\bibnamefont {Zheng}}, \bibinfo
  {author} {\bibfnamefont {S.}~\bibnamefont {Kowalski}}, \bibinfo {author}
  {\bibfnamefont {M.}~\bibnamefont {Barbui}}, \bibinfo {author} {\bibfnamefont
  {M.~R.~D.}\ \bibnamefont {Rodrigues}}, \bibinfo {author} {\bibfnamefont
  {K.}~\bibnamefont {Schmidt}}, \bibinfo {author} {\bibfnamefont
  {D.}~\bibnamefont {Fabris}}, \bibinfo {author} {\bibfnamefont
  {M.}~\bibnamefont {Lunardon}}, \bibinfo {author} {\bibfnamefont
  {S.}~\bibnamefont {Moretto}}, \bibinfo {author} {\bibfnamefont
  {G.}~\bibnamefont {Nebbia}}, \bibinfo {author} {\bibfnamefont
  {S.}~\bibnamefont {Pesente}}, \bibinfo {author} {\bibfnamefont
  {V.}~\bibnamefont {Rizzi}}, \bibinfo {author} {\bibfnamefont
  {G.}~\bibnamefont {Viesti}}, \bibinfo {author} {\bibfnamefont
  {M.}~\bibnamefont {Cinausero}}, \bibinfo {author} {\bibfnamefont
  {G.}~\bibnamefont {Prete}}, \bibinfo {author} {\bibfnamefont
  {T.}~\bibnamefont {Keutgen}}, \bibinfo {author} {\bibfnamefont
  {Y.}~\bibnamefont {El~Masri}}, \bibinfo {author} {\bibfnamefont
  {Z.}~\bibnamefont {Majka}}, \ and\ \bibinfo {author} {\bibfnamefont {Y.~G.}\
  \bibnamefont {Ma}},\ }\href {\doibase 10.1103/PhysRevLett.108.172701}
  {\bibfield  {journal} {\bibinfo  {journal} {Phys. Rev. Lett.}\ }\textbf
  {\bibinfo {volume} {108}},\ \bibinfo {pages} {172701} (\bibinfo {year}
  {2012})}\BibitemShut {NoStop}%
\bibitem [{\citenamefont {Bougault}\ \emph {et~al.}(2020)\citenamefont
  {Bougault}, \citenamefont {Bonnet}, \citenamefont {Borderie}, \citenamefont
  {Chbihi}, \citenamefont {Frankland}, \citenamefont {Galichet}, \citenamefont
  {Gruyer}, \citenamefont {Henri}, \citenamefont {Commara}, \citenamefont
  {Neindre}, \citenamefont {Lombardo}, \citenamefont {Lopez}, \citenamefont
  {Manduci}, \citenamefont {Parl\^{o}g}, \citenamefont {Roy}, \citenamefont
  {Verde},\ and\ \citenamefont {Vigilante}}]{Bougault2020_JPG47-025103}%
  \BibitemOpen
  \bibfield  {author} {\bibinfo {author} {\bibfnamefont {R.}~\bibnamefont
  {Bougault}}, \bibinfo {author} {\bibfnamefont {E.}~\bibnamefont {Bonnet}},
  \bibinfo {author} {\bibfnamefont {B.}~\bibnamefont {Borderie}}, \bibinfo
  {author} {\bibfnamefont {A.}~\bibnamefont {Chbihi}}, \bibinfo {author}
  {\bibfnamefont {J.~D.}\ \bibnamefont {Frankland}}, \bibinfo {author}
  {\bibfnamefont {E.}~\bibnamefont {Galichet}}, \bibinfo {author}
  {\bibfnamefont {D.}~\bibnamefont {Gruyer}}, \bibinfo {author} {\bibfnamefont
  {M.}~\bibnamefont {Henri}}, \bibinfo {author} {\bibfnamefont {M.~L.}\
  \bibnamefont {Commara}}, \bibinfo {author} {\bibfnamefont {N.~L.}\
  \bibnamefont {Neindre}}, \bibinfo {author} {\bibfnamefont {I.}~\bibnamefont
  {Lombardo}}, \bibinfo {author} {\bibfnamefont {O.}~\bibnamefont {Lopez}},
  \bibinfo {author} {\bibfnamefont {L.}~\bibnamefont {Manduci}}, \bibinfo
  {author} {\bibfnamefont {M.}~\bibnamefont {Parl\^{o}g}}, \bibinfo {author}
  {\bibfnamefont {R.}~\bibnamefont {Roy}}, \bibinfo {author} {\bibfnamefont
  {G.}~\bibnamefont {Verde}}, \ and\ \bibinfo {author} {\bibfnamefont
  {M.}~\bibnamefont {Vigilante}},\ }\href {\doibase 10.1088/1361-6471/ab56ba}
  {\bibfield  {journal} {\bibinfo  {journal} {J. Phys. G: Nucl. Part. Phys.}\
  }\textbf {\bibinfo {volume} {47}},\ \bibinfo {pages} {025103} (\bibinfo
  {year} {2020})}\BibitemShut {NoStop}%
\bibitem [{\citenamefont {Cust\'odio}\ \emph {et~al.}(2025)\citenamefont
  {Cust\'odio}, \citenamefont {Rebillard-Souli\'e}, \citenamefont {Bougault},
  \citenamefont {Gruyer}, \citenamefont {Gulminelli}, \citenamefont {Malik},
  \citenamefont {Pais},\ and\ \citenamefont
  {Provid\^encia}}]{Custodio2025_PRL134-082304}%
  \BibitemOpen
  \bibfield  {author} {\bibinfo {author} {\bibfnamefont {T.}~\bibnamefont
  {Cust\'odio}}, \bibinfo {author} {\bibfnamefont {A.}~\bibnamefont
  {Rebillard-Souli\'e}}, \bibinfo {author} {\bibfnamefont {R.}~\bibnamefont
  {Bougault}}, \bibinfo {author} {\bibfnamefont {D.}~\bibnamefont {Gruyer}},
  \bibinfo {author} {\bibfnamefont {F.}~\bibnamefont {Gulminelli}}, \bibinfo
  {author} {\bibfnamefont {T.}~\bibnamefont {Malik}}, \bibinfo {author}
  {\bibfnamefont {H.}~\bibnamefont {Pais}}, \ and\ \bibinfo {author}
  {\bibfnamefont {C.~m.~c.}\ \bibnamefont {Provid\^encia}},\ }\href {\doibase
  10.1103/PhysRevLett.134.082304} {\bibfield  {journal} {\bibinfo  {journal}
  {Phys. Rev. Lett.}\ }\textbf {\bibinfo {volume} {134}},\ \bibinfo {pages}
  {082304} (\bibinfo {year} {2025})}\BibitemShut {NoStop}%
\bibitem [{\citenamefont {Arcones}\ \emph {et~al.}(2008)\citenamefont
  {Arcones}, \citenamefont {Mart\'{\i}nez-Pinedo}, \citenamefont {O'Connor},
  \citenamefont {Schwenk}, \citenamefont {Janka}, \citenamefont {Horowitz},\
  and\ \citenamefont {Langanke}}]{Arcones2008_PRC78-015806}%
  \BibitemOpen
  \bibfield  {author} {\bibinfo {author} {\bibfnamefont {A.}~\bibnamefont
  {Arcones}}, \bibinfo {author} {\bibfnamefont {G.}~\bibnamefont
  {Mart\'{\i}nez-Pinedo}}, \bibinfo {author} {\bibfnamefont {E.}~\bibnamefont
  {O'Connor}}, \bibinfo {author} {\bibfnamefont {A.}~\bibnamefont {Schwenk}},
  \bibinfo {author} {\bibfnamefont {H.-T.}\ \bibnamefont {Janka}}, \bibinfo
  {author} {\bibfnamefont {C.~J.}\ \bibnamefont {Horowitz}}, \ and\ \bibinfo
  {author} {\bibfnamefont {K.}~\bibnamefont {Langanke}},\ }\href {\doibase
  10.1103/PhysRevC.78.015806} {\bibfield  {journal} {\bibinfo  {journal} {Phys.
  Rev. C}\ }\textbf {\bibinfo {volume} {78}},\ \bibinfo {pages} {015806}
  (\bibinfo {year} {2008})}\BibitemShut {NoStop}%
\bibitem [{\citenamefont {Fischer}\ \emph {et~al.}(2020)\citenamefont
  {Fischer}, \citenamefont {Typel}, \citenamefont {R\"opke}, \citenamefont
  {Bastian},\ and\ \citenamefont
  {Mart\'{\i}nez-Pinedo}}]{Fischer2020_PRC102-055807}%
  \BibitemOpen
  \bibfield  {author} {\bibinfo {author} {\bibfnamefont {T.}~\bibnamefont
  {Fischer}}, \bibinfo {author} {\bibfnamefont {S.}~\bibnamefont {Typel}},
  \bibinfo {author} {\bibfnamefont {G.}~\bibnamefont {R\"opke}}, \bibinfo
  {author} {\bibfnamefont {N.-U.~F.}\ \bibnamefont {Bastian}}, \ and\ \bibinfo
  {author} {\bibfnamefont {G.}~\bibnamefont {Mart\'{\i}nez-Pinedo}},\ }\href
  {\doibase 10.1103/PhysRevC.102.055807} {\bibfield  {journal} {\bibinfo
  {journal} {Phys. Rev. C}\ }\textbf {\bibinfo {volume} {102}},\ \bibinfo
  {pages} {055807} (\bibinfo {year} {2020})}\BibitemShut {NoStop}%
\bibitem [{\citenamefont {Rosswog}(2015)}]{Rosswog2015_IJMPD24-1530012}%
  \BibitemOpen
  \bibfield  {author} {\bibinfo {author} {\bibfnamefont {S.}~\bibnamefont
  {Rosswog}},\ }\href {\doibase 10.1142/S0218271815300128} {\bibfield
  {journal} {\bibinfo  {journal} {Int. J. Mod. Phys. D}\ }\textbf {\bibinfo
  {volume} {24}},\ \bibinfo {pages} {1530012} (\bibinfo {year}
  {2015})}\BibitemShut {NoStop}%
\bibitem [{\citenamefont {Alford}\ \emph {et~al.}(2018)\citenamefont {Alford},
  \citenamefont {Bovard}, \citenamefont {Hanauske}, \citenamefont {Rezzolla},\
  and\ \citenamefont {Schwenzer}}]{Alford2018_PRL120-041101}%
  \BibitemOpen
  \bibfield  {author} {\bibinfo {author} {\bibfnamefont {M.~G.}\ \bibnamefont
  {Alford}}, \bibinfo {author} {\bibfnamefont {L.}~\bibnamefont {Bovard}},
  \bibinfo {author} {\bibfnamefont {M.}~\bibnamefont {Hanauske}}, \bibinfo
  {author} {\bibfnamefont {L.}~\bibnamefont {Rezzolla}}, \ and\ \bibinfo
  {author} {\bibfnamefont {K.}~\bibnamefont {Schwenzer}},\ }\href {\doibase
  10.1103/PhysRevLett.120.041101} {\bibfield  {journal} {\bibinfo  {journal}
  {Phys. Rev. Lett.}\ }\textbf {\bibinfo {volume} {120}},\ \bibinfo {pages}
  {041101} (\bibinfo {year} {2018})}\BibitemShut {NoStop}%
\bibitem [{\citenamefont {Fujibayashi}\ \emph {et~al.}(8 06)\citenamefont
  {Fujibayashi}, \citenamefont {Kiuchi}, \citenamefont {Nishimura},
  \citenamefont {Sekiguchi},\ and\ \citenamefont
  {Shibata}}]{Fujibayashi2018_ApJ860-64}%
  \BibitemOpen
  \bibfield  {author} {\bibinfo {author} {\bibfnamefont {S.}~\bibnamefont
  {Fujibayashi}}, \bibinfo {author} {\bibfnamefont {K.}~\bibnamefont {Kiuchi}},
  \bibinfo {author} {\bibfnamefont {N.}~\bibnamefont {Nishimura}}, \bibinfo
  {author} {\bibfnamefont {Y.}~\bibnamefont {Sekiguchi}}, \ and\ \bibinfo
  {author} {\bibfnamefont {M.}~\bibnamefont {Shibata}},\ }\href {\doibase
  10.3847/1538-4357/aabafd} {\bibfield  {journal} {\bibinfo  {journal}
  {Astrophys. J.}\ }\textbf {\bibinfo {volume} {860}},\ \bibinfo {pages} {64}
  (\bibinfo {year} {2018-06})}\BibitemShut {NoStop}%
\bibitem [{\citenamefont {{Meyer}}(1994)}]{Meyer1994_ARAA32-153}%
  \BibitemOpen
  \bibfield  {author} {\bibinfo {author} {\bibfnamefont {B.~S.}\ \bibnamefont
  {{Meyer}}},\ }\href {\doibase 10.1146/annurev.aa.32.090194.001101} {\bibfield
   {journal} {\bibinfo  {journal} {Annu. Rev. Astron. Astrophys.}\ }\textbf
  {\bibinfo {volume} {32}},\ \bibinfo {pages} {153} (\bibinfo {year}
  {1994})}\BibitemShut {NoStop}%
\bibitem [{\citenamefont {R\"opke}\ \emph {et~al.}(1982)\citenamefont
  {R\"opke}, \citenamefont {M\"unchow},\ and\ \citenamefont
  {Schulz}}]{Roepke1982_NPA379-536}%
  \BibitemOpen
  \bibfield  {author} {\bibinfo {author} {\bibfnamefont {G.}~\bibnamefont
  {R\"opke}}, \bibinfo {author} {\bibfnamefont {L.}~\bibnamefont {M\"unchow}},
  \ and\ \bibinfo {author} {\bibfnamefont {H.}~\bibnamefont {Schulz}},\ }\href
  {\doibase https://doi.org/10.1016/0375-9474(82)90013-6} {\bibfield  {journal}
  {\bibinfo  {journal} {Nucl. Phys. A}\ }\textbf {\bibinfo {volume} {379}},\
  \bibinfo {pages} {536} (\bibinfo {year} {1982})}\BibitemShut {NoStop}%
\bibitem [{\citenamefont {Horowitz}\ and\ \citenamefont
  {Schwenk}(2006)}]{Horowitz2006_NPA776-55}%
  \BibitemOpen
  \bibfield  {author} {\bibinfo {author} {\bibfnamefont {C.}~\bibnamefont
  {Horowitz}}\ and\ \bibinfo {author} {\bibfnamefont {A.}~\bibnamefont
  {Schwenk}},\ }\href {\doibase
  https://doi.org/10.1016/j.nuclphysa.2006.05.009} {\bibfield  {journal}
  {\bibinfo  {journal} {Nucl. Phys. A}\ }\textbf {\bibinfo {volume} {776}},\
  \bibinfo {pages} {55} (\bibinfo {year} {2006})}\BibitemShut {NoStop}%
\bibitem [{\citenamefont {Lattimer}\ and\ \citenamefont {{Douglas
  Swesty}}(1991)}]{Lattimer1991_NPA535-331}%
  \BibitemOpen
  \bibfield  {author} {\bibinfo {author} {\bibfnamefont {J.~M.}\ \bibnamefont
  {Lattimer}}\ and\ \bibinfo {author} {\bibfnamefont {F.}~\bibnamefont
  {{Douglas Swesty}}},\ }\href {\doibase
  https://doi.org/10.1016/0375-9474(91)90452-C} {\bibfield  {journal} {\bibinfo
   {journal} {Nucl. Phys. A}\ }\textbf {\bibinfo {volume} {535}},\ \bibinfo
  {pages} {331} (\bibinfo {year} {1991})}\BibitemShut {NoStop}%
\bibitem [{\citenamefont {Shen}\ \emph {et~al.}(2011)\citenamefont {Shen},
  \citenamefont {Toki}, \citenamefont {Oyamatsu},\ and\ \citenamefont
  {Sumiyoshi}}]{Shen2011_ApJ197-20}%
  \BibitemOpen
  \bibfield  {author} {\bibinfo {author} {\bibfnamefont {H.}~\bibnamefont
  {Shen}}, \bibinfo {author} {\bibfnamefont {H.}~\bibnamefont {Toki}}, \bibinfo
  {author} {\bibfnamefont {K.}~\bibnamefont {Oyamatsu}}, \ and\ \bibinfo
  {author} {\bibfnamefont {K.}~\bibnamefont {Sumiyoshi}},\ }\href
  {http://stacks.iop.org/0067-0049/197/i=2/a=20} {\bibfield  {journal}
  {\bibinfo  {journal} {Astrophys. J.}\ }\textbf {\bibinfo {volume} {197}},\
  \bibinfo {pages} {20} (\bibinfo {year} {2011})}\BibitemShut {NoStop}%
\bibitem [{\citenamefont {Hempel}\ and\ \citenamefont
  {Schaffner-Bielich}(2010)}]{Hempel2010_NPA837-210}%
  \BibitemOpen
  \bibfield  {author} {\bibinfo {author} {\bibfnamefont {M.}~\bibnamefont
  {Hempel}}\ and\ \bibinfo {author} {\bibfnamefont {J.}~\bibnamefont
  {Schaffner-Bielich}},\ }\href {\doibase
  https://doi.org/10.1016/j.nuclphysa.2010.02.010} {\bibfield  {journal}
  {\bibinfo  {journal} {Nucl. Phys. A}\ }\textbf {\bibinfo {volume} {837}},\
  \bibinfo {pages} {210 } (\bibinfo {year} {2010})}\BibitemShut {NoStop}%
\bibitem [{\citenamefont {Raduta}\ and\ \citenamefont
  {Gulminelli}(2010)}]{Raduta2010_PRC82-065801}%
  \BibitemOpen
  \bibfield  {author} {\bibinfo {author} {\bibfnamefont {A.~R.}\ \bibnamefont
  {Raduta}}\ and\ \bibinfo {author} {\bibfnamefont {F.}~\bibnamefont
  {Gulminelli}},\ }\href {\doibase 10.1103/PhysRevC.82.065801} {\bibfield
  {journal} {\bibinfo  {journal} {Phys. Rev. C}\ }\textbf {\bibinfo {volume}
  {82}},\ \bibinfo {pages} {065801} (\bibinfo {year} {2010})}\BibitemShut
  {NoStop}%
\bibitem [{\citenamefont {Typel}\ \emph {et~al.}(2010)\citenamefont {Typel},
  \citenamefont {R\"opke}, \citenamefont {Kl\"ahn}, \citenamefont {Blaschke},\
  and\ \citenamefont {Wolter}}]{Typel2010_PRC81-015803}%
  \BibitemOpen
  \bibfield  {author} {\bibinfo {author} {\bibfnamefont {S.}~\bibnamefont
  {Typel}}, \bibinfo {author} {\bibfnamefont {G.}~\bibnamefont {R\"opke}},
  \bibinfo {author} {\bibfnamefont {T.}~\bibnamefont {Kl\"ahn}}, \bibinfo
  {author} {\bibfnamefont {D.}~\bibnamefont {Blaschke}}, \ and\ \bibinfo
  {author} {\bibfnamefont {H.~H.}\ \bibnamefont {Wolter}},\ }\href {\doibase
  10.1103/PhysRevC.81.015803} {\bibfield  {journal} {\bibinfo  {journal} {Phys.
  Rev. C}\ }\textbf {\bibinfo {volume} {81}},\ \bibinfo {pages} {015803}
  (\bibinfo {year} {2010})}\BibitemShut {NoStop}%
\bibitem [{\citenamefont {R\"opke}(2014)}]{Roepke2014_JP569-012031}%
  \BibitemOpen
  \bibfield  {author} {\bibinfo {author} {\bibfnamefont {G.}~\bibnamefont
  {R\"opke}},\ }\href {http://stacks.iop.org/1742-6596/569/i=1/a=012031}
  {\bibfield  {journal} {\bibinfo  {journal} {J. Phys.: Conf. Ser.}\ }\textbf
  {\bibinfo {volume} {569}},\ \bibinfo {pages} {012031} (\bibinfo {year}
  {2014})}\BibitemShut {NoStop}%
\bibitem [{\citenamefont {R\"opke}(2020)}]{Roepke2020_PRC101-064310}%
  \BibitemOpen
  \bibfield  {author} {\bibinfo {author} {\bibfnamefont {G.}~\bibnamefont
  {R\"opke}},\ }\href {\doibase 10.1103/PhysRevC.101.064310} {\bibfield
  {journal} {\bibinfo  {journal} {Phys. Rev. C}\ }\textbf {\bibinfo {volume}
  {101}},\ \bibinfo {pages} {064310} (\bibinfo {year} {2020})}\BibitemShut
  {NoStop}%
\bibitem [{\citenamefont {Avancini}\ \emph {et~al.}(2010)\citenamefont
  {Avancini}, \citenamefont {Barros}, \citenamefont {Menezes},\ and\
  \citenamefont {Provid\^encia}}]{Avancini2010_PRC82-025808}%
  \BibitemOpen
  \bibfield  {author} {\bibinfo {author} {\bibfnamefont {S.~S.}\ \bibnamefont
  {Avancini}}, \bibinfo {author} {\bibfnamefont {C.~C.}\ \bibnamefont
  {Barros}}, \bibinfo {author} {\bibfnamefont {D.~P.}\ \bibnamefont {Menezes}},
  \ and\ \bibinfo {author} {\bibfnamefont {C.}~\bibnamefont {Provid\^encia}},\
  }\href {\doibase 10.1103/PhysRevC.82.025808} {\bibfield  {journal} {\bibinfo
  {journal} {Phys. Rev. C}\ }\textbf {\bibinfo {volume} {82}},\ \bibinfo
  {pages} {025808} (\bibinfo {year} {2010})}\BibitemShut {NoStop}%
\bibitem [{\citenamefont {Ferreira}\ and\ \citenamefont
  {Provid\^encia}(2012)}]{Ferreira2012_PRC85-055811}%
  \BibitemOpen
  \bibfield  {author} {\bibinfo {author} {\bibfnamefont {M.}~\bibnamefont
  {Ferreira}}\ and\ \bibinfo {author} {\bibfnamefont {C.~m.~c.}\ \bibnamefont
  {Provid\^encia}},\ }\href {\doibase 10.1103/PhysRevC.85.055811} {\bibfield
  {journal} {\bibinfo  {journal} {Phys. Rev. C}\ }\textbf {\bibinfo {volume}
  {85}},\ \bibinfo {pages} {055811} (\bibinfo {year} {2012})}\BibitemShut
  {NoStop}%
\bibitem [{\citenamefont {Brockmann}\ and\ \citenamefont
  {Weise}(1977)}]{Brockmann1977_PLB69-167}%
  \BibitemOpen
  \bibfield  {author} {\bibinfo {author} {\bibfnamefont {R.}~\bibnamefont
  {Brockmann}}\ and\ \bibinfo {author} {\bibfnamefont {W.}~\bibnamefont
  {Weise}},\ }\href {\doibase http://dx.doi.org/10.1016/0370-2693(77)90635-9}
  {\bibfield  {journal} {\bibinfo  {journal} {Phys. Lett. B}\ }\textbf
  {\bibinfo {volume} {69}},\ \bibinfo {pages} {167} (\bibinfo {year}
  {1977})}\BibitemShut {NoStop}%
\bibitem [{\citenamefont {Boguta}\ and\ \citenamefont
  {Bohrmann}(1981)}]{Boguta1981_PLB102-93}%
  \BibitemOpen
  \bibfield  {author} {\bibinfo {author} {\bibfnamefont {J.}~\bibnamefont
  {Boguta}}\ and\ \bibinfo {author} {\bibfnamefont {S.}~\bibnamefont
  {Bohrmann}},\ }\href {\doibase
  http://dx.doi.org/10.1016/0370-2693(81)91037-6} {\bibfield  {journal}
  {\bibinfo  {journal} {Phys. Lett. B}\ }\textbf {\bibinfo {volume} {102}},\
  \bibinfo {pages} {93} (\bibinfo {year} {1981})}\BibitemShut {NoStop}%
\bibitem [{\citenamefont {Mare{\v{s}}}\ and\ \citenamefont
  {{\v{Z}}ofka}(1989)}]{Mares1989_ZPA333-209}%
  \BibitemOpen
  \bibfield  {author} {\bibinfo {author} {\bibfnamefont {J.}~\bibnamefont
  {Mare{\v{s}}}}\ and\ \bibinfo {author} {\bibfnamefont {J.}~\bibnamefont
  {{\v{Z}}ofka}},\ }\href {\doibase 10.1007/BF01565152} {\bibfield  {journal}
  {\bibinfo  {journal} {Z. Phys. A}\ }\textbf {\bibinfo {volume} {333}},\
  \bibinfo {pages} {209} (\bibinfo {year} {1989})}\BibitemShut {NoStop}%
\bibitem [{\citenamefont {Reinhard}(1989)}]{Reinhard1989_RPP52-439}%
  \BibitemOpen
  \bibfield  {author} {\bibinfo {author} {\bibfnamefont {P.-G.}\ \bibnamefont
  {Reinhard}},\ }\href {http://stacks.iop.org/0034-4885/52/i=4/a=002}
  {\bibfield  {journal} {\bibinfo  {journal} {Rep. Prog. Phys.}\ }\textbf
  {\bibinfo {volume} {52}},\ \bibinfo {pages} {439} (\bibinfo {year}
  {1989})}\BibitemShut {NoStop}%
\bibitem [{\citenamefont {Sugahara}\ and\ \citenamefont
  {Toki}(1994{\natexlab{a}})}]{Toki1994_PTP92-803}%
  \BibitemOpen
  \bibfield  {author} {\bibinfo {author} {\bibfnamefont {Y.}~\bibnamefont
  {Sugahara}}\ and\ \bibinfo {author} {\bibfnamefont {H.}~\bibnamefont
  {Toki}},\ }\href {\doibase 10.1143/ptp/92.4.803} {\bibfield  {journal}
  {\bibinfo  {journal} {Prog. Theor. Phys.}\ }\textbf {\bibinfo {volume}
  {92}},\ \bibinfo {pages} {803} (\bibinfo {year}
  {1994}{\natexlab{a}})}\BibitemShut {NoStop}%
\bibitem [{\citenamefont {Ring}(1996)}]{Ring1996_PPNP37_193-263}%
  \BibitemOpen
  \bibfield  {author} {\bibinfo {author} {\bibfnamefont {P.}~\bibnamefont
  {Ring}},\ }\href {\doibase 10.1016/0146-6410(96)00054-3} {\bibfield
  {journal} {\bibinfo  {journal} {Prog. Part. Nucl. Phys.}\ }\textbf {\bibinfo
  {volume} {37}},\ \bibinfo {pages} {193} (\bibinfo {year} {1996})}\BibitemShut
  {NoStop}%
\bibitem [{\citenamefont {Meng}\ \emph {et~al.}(2006)\citenamefont {Meng},
  \citenamefont {Toki}, \citenamefont {Zhou}, \citenamefont {Zhang},
  \citenamefont {Long},\ and\ \citenamefont {Geng}}]{Meng2006_PPNP57-470}%
  \BibitemOpen
  \bibfield  {author} {\bibinfo {author} {\bibfnamefont {J.}~\bibnamefont
  {Meng}}, \bibinfo {author} {\bibfnamefont {H.}~\bibnamefont {Toki}}, \bibinfo
  {author} {\bibfnamefont {S.}~\bibnamefont {Zhou}}, \bibinfo {author}
  {\bibfnamefont {S.}~\bibnamefont {Zhang}}, \bibinfo {author} {\bibfnamefont
  {W.}~\bibnamefont {Long}}, \ and\ \bibinfo {author} {\bibfnamefont
  {L.}~\bibnamefont {Geng}},\ }\href {\doibase 10.1016/j.ppnp.2005.06.001}
  {\bibfield  {journal} {\bibinfo  {journal} {Prog. Part. Nucl. Phys.}\
  }\textbf {\bibinfo {volume} {57}},\ \bibinfo {pages} {470} (\bibinfo {year}
  {2006})}\BibitemShut {NoStop}%
\bibitem [{\citenamefont {Paar}\ \emph {et~al.}(2007)\citenamefont {Paar},
  \citenamefont {Vretenar}, \citenamefont {Khan},\ and\ \citenamefont
  {Col\`{o}}}]{Paar2007_RPP70-691}%
  \BibitemOpen
  \bibfield  {author} {\bibinfo {author} {\bibfnamefont {N.}~\bibnamefont
  {Paar}}, \bibinfo {author} {\bibfnamefont {D.}~\bibnamefont {Vretenar}},
  \bibinfo {author} {\bibfnamefont {E.}~\bibnamefont {Khan}}, \ and\ \bibinfo
  {author} {\bibfnamefont {G.}~\bibnamefont {Col\`{o}}},\ }\href
  {http://stacks.iop.org/0034-4885/70/i=5/a=R02} {\bibfield  {journal}
  {\bibinfo  {journal} {Rep. Prog. Phys.}\ }\textbf {\bibinfo {volume} {70}},\
  \bibinfo {pages} {691} (\bibinfo {year} {2007})}\BibitemShut {NoStop}%
\bibitem [{\citenamefont {Tanimura}\ and\ \citenamefont
  {Hagino}(2012)}]{Tanimura2012_PRC85-014306}%
  \BibitemOpen
  \bibfield  {author} {\bibinfo {author} {\bibfnamefont {Y.}~\bibnamefont
  {Tanimura}}\ and\ \bibinfo {author} {\bibfnamefont {K.}~\bibnamefont
  {Hagino}},\ }\href {\doibase 10.1103/PhysRevC.85.014306} {\bibfield
  {journal} {\bibinfo  {journal} {Phys. Rev. C}\ }\textbf {\bibinfo {volume}
  {85}},\ \bibinfo {pages} {014306} (\bibinfo {year} {2012})}\BibitemShut
  {NoStop}%
\bibitem [{\citenamefont {WANG}\ \emph {et~al.}(2013)\citenamefont {WANG},
  \citenamefont {SANG}, \citenamefont {WANG},\ and\ \citenamefont
  {LV}}]{Wang2013_CTP60-479}%
  \BibitemOpen
  \bibfield  {author} {\bibinfo {author} {\bibfnamefont {X.-S.}\ \bibnamefont
  {WANG}}, \bibinfo {author} {\bibfnamefont {H.-Y.}\ \bibnamefont {SANG}},
  \bibinfo {author} {\bibfnamefont {J.-H.}\ \bibnamefont {WANG}}, \ and\
  \bibinfo {author} {\bibfnamefont {H.-F.}\ \bibnamefont {LV}},\ }\href
  {http://ctp.itp.ac.cn/EN/abstract/abstract16149.shtml#} {\bibfield  {journal}
  {\bibinfo  {journal} {Commun. Theor. Phys.}\ }\textbf {\bibinfo {volume}
  {60}},\ \bibinfo {pages} {479} (\bibinfo {year} {2013})}\BibitemShut
  {NoStop}%
\bibitem [{\citenamefont {Meng}\ and\ \citenamefont
  {Zhou}(2015)}]{Meng2015_JPG42-093101}%
  \BibitemOpen
  \bibfield  {author} {\bibinfo {author} {\bibfnamefont {J.}~\bibnamefont
  {Meng}}\ and\ \bibinfo {author} {\bibfnamefont {S.~G.}\ \bibnamefont
  {Zhou}},\ }\href {http://stacks.iop.org/0954-3899/42/i=9/a=093101} {\bibfield
   {journal} {\bibinfo  {journal} {J. Phys. G: Nucl. Part. Phys.}\ }\textbf
  {\bibinfo {volume} {42}},\ \bibinfo {pages} {093101} (\bibinfo {year}
  {2015})}\BibitemShut {NoStop}%
\bibitem [{\citenamefont {Meng}(2016)}]{Meng2016_RDFNS}%
  \BibitemOpen
  \bibinfo {editor} {\bibfnamefont {J.}~\bibnamefont {Meng}},\ ed.,\ \href
  {\doibase 10.1142/9872} {\emph {\bibinfo {title} {Relativistic Density
  Functional for Nuclear Structure}}},\ \bibinfo {series} {International Review
  of Nuclear Physics}, Vol.~\bibinfo {volume} {10}\ (\bibinfo  {publisher}
  {World Scientific Pub Co Pte Lt},\ \bibinfo {year} {2016})\BibitemShut
  {NoStop}%
\bibitem [{\citenamefont {Chen}\ \emph {et~al.}(2021)\citenamefont {Chen},
  \citenamefont {Sun}, \citenamefont {Li},\ and\ \citenamefont
  {Sun}}]{Chen2021_SCPMA64-282011}%
  \BibitemOpen
  \bibfield  {author} {\bibinfo {author} {\bibfnamefont {C.}~\bibnamefont
  {Chen}}, \bibinfo {author} {\bibfnamefont {Q.-K.}\ \bibnamefont {Sun}},
  \bibinfo {author} {\bibfnamefont {Y.-X.}\ \bibnamefont {Li}}, \ and\ \bibinfo
  {author} {\bibfnamefont {T.-T.}\ \bibnamefont {Sun}},\ }\href {\doibase
  10.1007/s11433-021-1721-1} {\bibfield  {journal} {\bibinfo  {journal} {Sci.
  China Phys. Mech. Astron.}\ }\textbf {\bibinfo {volume} {64}},\ \bibinfo
  {pages} {282011} (\bibinfo {year} {2021})}\BibitemShut {NoStop}%
\bibitem [{\citenamefont {Typel}\ and\ \citenamefont
  {Wolter}(1999)}]{Typel1999_NPA656-331}%
  \BibitemOpen
  \bibfield  {author} {\bibinfo {author} {\bibfnamefont {S.}~\bibnamefont
  {Typel}}\ and\ \bibinfo {author} {\bibfnamefont {H.}~\bibnamefont {Wolter}},\
  }\href {\doibase http://dx.doi.org/10.1016/S0375-9474(99)00310-3} {\bibfield
  {journal} {\bibinfo  {journal} {Nucl. Phys. A}\ }\textbf {\bibinfo {volume}
  {656}},\ \bibinfo {pages} {331} (\bibinfo {year} {1999})}\BibitemShut
  {NoStop}%
\bibitem [{\citenamefont {Vretenar}\ \emph {et~al.}(1998)\citenamefont
  {Vretenar}, \citenamefont {P\"oschl}, \citenamefont {Lalazissis},\ and\
  \citenamefont {Ring}}]{Vretenar1998_PRC57-R1060}%
  \BibitemOpen
  \bibfield  {author} {\bibinfo {author} {\bibfnamefont {D.}~\bibnamefont
  {Vretenar}}, \bibinfo {author} {\bibfnamefont {W.}~\bibnamefont {P\"oschl}},
  \bibinfo {author} {\bibfnamefont {G.~A.}\ \bibnamefont {Lalazissis}}, \ and\
  \bibinfo {author} {\bibfnamefont {P.}~\bibnamefont {Ring}},\ }\href {\doibase
  10.1103/PhysRevC.57.R1060} {\bibfield  {journal} {\bibinfo  {journal} {Phys.
  Rev. C}\ }\textbf {\bibinfo {volume} {57}},\ \bibinfo {pages} {R1060}
  (\bibinfo {year} {1998})}\BibitemShut {NoStop}%
\bibitem [{\citenamefont {Lu}\ \emph {et~al.}(2011)\citenamefont {Lu},
  \citenamefont {Zhao},\ and\ \citenamefont {Zhou}}]{Lu2011_PRC84-014328}%
  \BibitemOpen
  \bibfield  {author} {\bibinfo {author} {\bibfnamefont {B.-N.}\ \bibnamefont
  {Lu}}, \bibinfo {author} {\bibfnamefont {E.-G.}\ \bibnamefont {Zhao}}, \ and\
  \bibinfo {author} {\bibfnamefont {S.-G.}\ \bibnamefont {Zhou}},\ }\href
  {\doibase 10.1103/PhysRevC.84.014328} {\bibfield  {journal} {\bibinfo
  {journal} {Phys. Rev. C}\ }\textbf {\bibinfo {volume} {84}},\ \bibinfo
  {pages} {014328} (\bibinfo {year} {2011})}\BibitemShut {NoStop}%
\bibitem [{\citenamefont {Wei}\ \emph {et~al.}(2020)\citenamefont {Wei},
  \citenamefont {Zhao}, \citenamefont {Wang}, \citenamefont {Geng},
  \citenamefont {Sun}, \citenamefont {Niu},\ and\ \citenamefont
  {Long}}]{Wei2020_CPC44-074107}%
  \BibitemOpen
  \bibfield  {author} {\bibinfo {author} {\bibfnamefont {B.}~\bibnamefont
  {Wei}}, \bibinfo {author} {\bibfnamefont {Q.}~\bibnamefont {Zhao}}, \bibinfo
  {author} {\bibfnamefont {Z.-H.}\ \bibnamefont {Wang}}, \bibinfo {author}
  {\bibfnamefont {J.}~\bibnamefont {Geng}}, \bibinfo {author} {\bibfnamefont
  {B.-Y.}\ \bibnamefont {Sun}}, \bibinfo {author} {\bibfnamefont {Y.-F.}\
  \bibnamefont {Niu}}, \ and\ \bibinfo {author} {\bibfnamefont {W.-H.}\
  \bibnamefont {Long}},\ }\href {\doibase 10.1088/1674-1137/44/7/074107}
  {\bibfield  {journal} {\bibinfo  {journal} {Chin. Phys. C}\ }\textbf
  {\bibinfo {volume} {44}},\ \bibinfo {pages} {074107} (\bibinfo {year}
  {2020})}\BibitemShut {NoStop}%
\bibitem [{\citenamefont {Taninah}\ \emph {et~al.}(2020)\citenamefont
  {Taninah}, \citenamefont {Agbemava}, \citenamefont {Afanasjev},\ and\
  \citenamefont {Ring}}]{Taninah2020_PLB800-135065}%
  \BibitemOpen
  \bibfield  {author} {\bibinfo {author} {\bibfnamefont {A.}~\bibnamefont
  {Taninah}}, \bibinfo {author} {\bibfnamefont {S.}~\bibnamefont {Agbemava}},
  \bibinfo {author} {\bibfnamefont {A.}~\bibnamefont {Afanasjev}}, \ and\
  \bibinfo {author} {\bibfnamefont {P.}~\bibnamefont {Ring}},\ }\href {\doibase
  https://doi.org/10.1016/j.physletb.2019.135065} {\bibfield  {journal}
  {\bibinfo  {journal} {Phys. Lett. B}\ }\textbf {\bibinfo {volume} {800}},\
  \bibinfo {pages} {135065} (\bibinfo {year} {2020})}\BibitemShut {NoStop}%
\bibitem [{\citenamefont {Glendenning}(2000)}]{Glendenning2000}%
  \BibitemOpen
  \bibfield  {author} {\bibinfo {author} {\bibfnamefont {N.}~\bibnamefont
  {Glendenning}},\ }\href {http://www.springer.com/cn/book/9780387989778}
  {\emph {\bibinfo {title} {Compact Stars. Nuclear Physics, Particle Physics,
  and General Relativity}}},\ \bibinfo {edition} {2nd}\ ed.,\ ISBN
  978-0-387-98977-8\ (\bibinfo  {publisher} {Springer-Verlag},\ \bibinfo
  {address} {Berlin},\ \bibinfo {year} {2000})\BibitemShut {NoStop}%
\bibitem [{\citenamefont {Ban}\ \emph {et~al.}(2004)\citenamefont {Ban},
  \citenamefont {Li}, \citenamefont {Zhang}, \citenamefont {Jia}, \citenamefont
  {Sang},\ and\ \citenamefont {Meng}}]{Ban2004_PRC69-045805}%
  \BibitemOpen
  \bibfield  {author} {\bibinfo {author} {\bibfnamefont {S.~F.}\ \bibnamefont
  {Ban}}, \bibinfo {author} {\bibfnamefont {J.}~\bibnamefont {Li}}, \bibinfo
  {author} {\bibfnamefont {S.~Q.}\ \bibnamefont {Zhang}}, \bibinfo {author}
  {\bibfnamefont {H.~Y.}\ \bibnamefont {Jia}}, \bibinfo {author} {\bibfnamefont
  {J.~P.}\ \bibnamefont {Sang}}, \ and\ \bibinfo {author} {\bibfnamefont
  {J.}~\bibnamefont {Meng}},\ }\href {\doibase 10.1103/PhysRevC.69.045805}
  {\bibfield  {journal} {\bibinfo  {journal} {Phys. Rev. C}\ }\textbf {\bibinfo
  {volume} {69}},\ \bibinfo {pages} {045805} (\bibinfo {year}
  {2004})}\BibitemShut {NoStop}%
\bibitem [{\citenamefont {Weber}\ \emph {et~al.}(2007)\citenamefont {Weber},
  \citenamefont {Negreiros}, \citenamefont {Rosenfield},\ and\ \citenamefont
  {Stejner}}]{Weber2007_PPNP59-94}%
  \BibitemOpen
  \bibfield  {author} {\bibinfo {author} {\bibfnamefont {F.}~\bibnamefont
  {Weber}}, \bibinfo {author} {\bibfnamefont {R.}~\bibnamefont {Negreiros}},
  \bibinfo {author} {\bibfnamefont {P.}~\bibnamefont {Rosenfield}}, \ and\
  \bibinfo {author} {\bibfnamefont {M.}~\bibnamefont {Stejner}},\ }\href
  {\doibase http://dx.doi.org/10.1016/j.ppnp.2006.12.008} {\bibfield  {journal}
  {\bibinfo  {journal} {Prog. Part. Nucl. Phys.}\ }\textbf {\bibinfo {volume}
  {59}},\ \bibinfo {pages} {94 } (\bibinfo {year} {2007})}\BibitemShut
  {NoStop}%
\bibitem [{\citenamefont {Long}\ \emph {et~al.}(2012)\citenamefont {Long},
  \citenamefont {Sun}, \citenamefont {Hagino},\ and\ \citenamefont
  {Sagawa}}]{Long2012_PRC85-025806}%
  \BibitemOpen
  \bibfield  {author} {\bibinfo {author} {\bibfnamefont {W.~H.}\ \bibnamefont
  {Long}}, \bibinfo {author} {\bibfnamefont {B.~Y.}\ \bibnamefont {Sun}},
  \bibinfo {author} {\bibfnamefont {K.}~\bibnamefont {Hagino}}, \ and\ \bibinfo
  {author} {\bibfnamefont {H.}~\bibnamefont {Sagawa}},\ }\href {\doibase
  10.1103/PhysRevC.85.025806} {\bibfield  {journal} {\bibinfo  {journal} {Phys.
  Rev. C}\ }\textbf {\bibinfo {volume} {85}},\ \bibinfo {pages} {025806}
  (\bibinfo {year} {2012})}\BibitemShut {NoStop}%
\bibitem [{\citenamefont {Sun}\ \emph {et~al.}(2012)\citenamefont {Sun},
  \citenamefont {Sun},\ and\ \citenamefont {Meng}}]{Sun2012_PRC86-014305}%
  \BibitemOpen
  \bibfield  {author} {\bibinfo {author} {\bibfnamefont {T.~T.}\ \bibnamefont
  {Sun}}, \bibinfo {author} {\bibfnamefont {B.~Y.}\ \bibnamefont {Sun}}, \ and\
  \bibinfo {author} {\bibfnamefont {J.}~\bibnamefont {Meng}},\ }\href {\doibase
  10.1103/PhysRevC.86.014305} {\bibfield  {journal} {\bibinfo  {journal} {Phys.
  Rev. C}\ }\textbf {\bibinfo {volume} {86}},\ \bibinfo {pages} {014305}
  (\bibinfo {year} {2012})}\BibitemShut {NoStop}%
\bibitem [{\citenamefont {Wang}\ \emph {et~al.}(2014)\citenamefont {Wang},
  \citenamefont {Zhang},\ and\ \citenamefont {Dong}}]{Wang2014_PRC90-055801}%
  \BibitemOpen
  \bibfield  {author} {\bibinfo {author} {\bibfnamefont {S.}~\bibnamefont
  {Wang}}, \bibinfo {author} {\bibfnamefont {H.~F.}\ \bibnamefont {Zhang}}, \
  and\ \bibinfo {author} {\bibfnamefont {J.~M.}\ \bibnamefont {Dong}},\ }\href
  {\doibase 10.1103/PhysRevC.90.055801} {\bibfield  {journal} {\bibinfo
  {journal} {Phys. Rev. C}\ }\textbf {\bibinfo {volume} {90}},\ \bibinfo
  {pages} {055801} (\bibinfo {year} {2014})}\BibitemShut {NoStop}%
\bibitem [{\citenamefont {Fedoseew}\ and\ \citenamefont
  {Lenske}(2015)}]{Fedoseew2015_PRC91-034307}%
  \BibitemOpen
  \bibfield  {author} {\bibinfo {author} {\bibfnamefont {A.}~\bibnamefont
  {Fedoseew}}\ and\ \bibinfo {author} {\bibfnamefont {H.}~\bibnamefont
  {Lenske}},\ }\href {\doibase 10.1103/PhysRevC.91.034307} {\bibfield
  {journal} {\bibinfo  {journal} {Phys. Rev. C}\ }\textbf {\bibinfo {volume}
  {91}},\ \bibinfo {pages} {034307} (\bibinfo {year} {2015})}\BibitemShut
  {NoStop}%
\bibitem [{\citenamefont {Gao}\ \emph {et~al.}(2017)\citenamefont {Gao},
  \citenamefont {Wang}, \citenamefont {Shan}, \citenamefont {Li},\ and\
  \citenamefont {Wang}}]{Gao2017_ApJ849-19}%
  \BibitemOpen
  \bibfield  {author} {\bibinfo {author} {\bibfnamefont {Z.-F.}\ \bibnamefont
  {Gao}}, \bibinfo {author} {\bibfnamefont {N.}~\bibnamefont {Wang}}, \bibinfo
  {author} {\bibfnamefont {H.}~\bibnamefont {Shan}}, \bibinfo {author}
  {\bibfnamefont {X.-D.}\ \bibnamefont {Li}}, \ and\ \bibinfo {author}
  {\bibfnamefont {W.}~\bibnamefont {Wang}},\ }\href
  {http://stacks.iop.org/0004-637X/849/i=1/a=19} {\bibfield  {journal}
  {\bibinfo  {journal} {Astrophys. J.}\ }\textbf {\bibinfo {volume} {849}},\
  \bibinfo {pages} {19} (\bibinfo {year} {2017})}\BibitemShut {NoStop}%
\bibitem [{\citenamefont {{Negele}}\ and\ \citenamefont
  {{Vautherin}}(1973)}]{Negele1973_NPA207-298}%
  \BibitemOpen
  \bibfield  {author} {\bibinfo {author} {\bibfnamefont {J.~W.}\ \bibnamefont
  {{Negele}}}\ and\ \bibinfo {author} {\bibfnamefont {D.}~\bibnamefont
  {{Vautherin}}},\ }\href {\doibase 10.1016/0375-9474(73)90349-7} {\bibfield
  {journal} {\bibinfo  {journal} {Nucl. Phys. A}\ }\textbf {\bibinfo {volume}
  {207}},\ \bibinfo {pages} {298} (\bibinfo {year} {1973})}\BibitemShut
  {NoStop}%
\bibitem [{\citenamefont {Bender}\ \emph {et~al.}(2000)\citenamefont {Bender},
  \citenamefont {Rutz}, \citenamefont {Reinhard},\ and\ \citenamefont
  {Maruhn}}]{Bender2000_EPJA7-467}%
  \BibitemOpen
  \bibfield  {author} {\bibinfo {author} {\bibfnamefont {M.}~\bibnamefont
  {Bender}}, \bibinfo {author} {\bibfnamefont {K.}~\bibnamefont {Rutz}},
  \bibinfo {author} {\bibfnamefont {P.-G.}\ \bibnamefont {Reinhard}}, \ and\
  \bibinfo {author} {\bibfnamefont {J.}~\bibnamefont {Maruhn}},\ }\href
  {\doibase 10.1007/PL00013645} {\bibfield  {journal} {\bibinfo  {journal}
  {Eur. Phys. J. A}\ }\textbf {\bibinfo {volume} {7}},\ \bibinfo {pages} {467}
  (\bibinfo {year} {2000})}\BibitemShut {NoStop}%
\bibitem [{\citenamefont {Rong}(2023)}]{Rong2023_PRC108-054314}%
  \BibitemOpen
  \bibfield  {author} {\bibinfo {author} {\bibfnamefont {Y.-T.}\ \bibnamefont
  {Rong}},\ }\href {\doibase 10.1103/PhysRevC.108.054314} {\bibfield  {journal}
  {\bibinfo  {journal} {Phys. Rev. C}\ }\textbf {\bibinfo {volume} {108}},\
  \bibinfo {pages} {054314} (\bibinfo {year} {2023})}\BibitemShut {NoStop}%
\bibitem [{\citenamefont {Lalazissis}\ \emph {et~al.}(1997)\citenamefont
  {Lalazissis}, \citenamefont {K\"onig},\ and\ \citenamefont
  {Ring}}]{Lalazissis1997_PRC55-540}%
  \BibitemOpen
  \bibfield  {author} {\bibinfo {author} {\bibfnamefont {G.~A.}\ \bibnamefont
  {Lalazissis}}, \bibinfo {author} {\bibfnamefont {J.}~\bibnamefont {K\"onig}},
  \ and\ \bibinfo {author} {\bibfnamefont {P.}~\bibnamefont {Ring}},\ }\href
  {\doibase 10.1103/PhysRevC.55.540} {\bibfield  {journal} {\bibinfo  {journal}
  {Phys. Rev. C}\ }\textbf {\bibinfo {volume} {55}},\ \bibinfo {pages} {540}
  (\bibinfo {year} {1997})}\BibitemShut {NoStop}%
\bibitem [{\citenamefont {Long}\ \emph {et~al.}(2004)\citenamefont {Long},
  \citenamefont {Meng}, \citenamefont {Giai},\ and\ \citenamefont
  {Zhou}}]{Long2004_PRC69-034319}%
  \BibitemOpen
  \bibfield  {author} {\bibinfo {author} {\bibfnamefont {W.-H.}\ \bibnamefont
  {Long}}, \bibinfo {author} {\bibfnamefont {J.}~\bibnamefont {Meng}}, \bibinfo
  {author} {\bibfnamefont {N.~V.}\ \bibnamefont {Giai}}, \ and\ \bibinfo
  {author} {\bibfnamefont {S.-G.}\ \bibnamefont {Zhou}},\ }\href {\doibase
  10.1103/PhysRevC.69.034319} {\bibfield  {journal} {\bibinfo  {journal} {Phys.
  Rev. C}\ }\textbf {\bibinfo {volume} {69}},\ \bibinfo {pages} {034319}
  (\bibinfo {year} {2004})}\BibitemShut {NoStop}%
\bibitem [{\citenamefont {Sugahara}\ and\ \citenamefont
  {Toki}(1994{\natexlab{b}})}]{Sugahara1994_NPA579-557}%
  \BibitemOpen
  \bibfield  {author} {\bibinfo {author} {\bibfnamefont {Y.}~\bibnamefont
  {Sugahara}}\ and\ \bibinfo {author} {\bibfnamefont {H.}~\bibnamefont
  {Toki}},\ }\href {\doibase http://dx.doi.org/10.1016/0375-9474(94)90923-7}
  {\bibfield  {journal} {\bibinfo  {journal} {Nucl. Phys. A}\ }\textbf
  {\bibinfo {volume} {579}},\ \bibinfo {pages} {557} (\bibinfo {year}
  {1994}{\natexlab{b}})}\BibitemShut {NoStop}%
\bibitem [{\citenamefont {Lalazissis}\ \emph {et~al.}(2005)\citenamefont
  {Lalazissis}, \citenamefont {Nik\u{s}i\'{c}}, \citenamefont {Vretenar},\ and\
  \citenamefont {Ring}}]{Lalazissis2005_PRC71-024312}%
  \BibitemOpen
  \bibfield  {author} {\bibinfo {author} {\bibfnamefont {G.~A.}\ \bibnamefont
  {Lalazissis}}, \bibinfo {author} {\bibfnamefont {T.}~\bibnamefont
  {Nik\u{s}i\'{c}}}, \bibinfo {author} {\bibfnamefont {D.}~\bibnamefont
  {Vretenar}}, \ and\ \bibinfo {author} {\bibfnamefont {P.}~\bibnamefont
  {Ring}},\ }\href {\doibase 10.1103/PhysRevC.71.024312} {\bibfield  {journal}
  {\bibinfo  {journal} {Phys. Rev. C}\ }\textbf {\bibinfo {volume} {71}},\
  \bibinfo {pages} {024312} (\bibinfo {year} {2005})}\BibitemShut {NoStop}%
\bibitem [{\citenamefont {Lenske}\ and\ \citenamefont
  {Fuchs}(1995)}]{Lenske1995_PLB345-355}%
  \BibitemOpen
  \bibfield  {author} {\bibinfo {author} {\bibfnamefont {H.}~\bibnamefont
  {Lenske}}\ and\ \bibinfo {author} {\bibfnamefont {C.}~\bibnamefont {Fuchs}},\
  }\href {\doibase https://doi.org/10.1016/0370-2693(94)01664-X} {\bibfield
  {journal} {\bibinfo  {journal} {Phys. Lett. B}\ }\textbf {\bibinfo {volume}
  {345}},\ \bibinfo {pages} {355 } (\bibinfo {year} {1995})}\BibitemShut
  {NoStop}%
\bibitem [{\citenamefont {Xia}\ \emph {et~al.}(2021)\citenamefont {Xia},
  \citenamefont {Maruyama}, \citenamefont {Yasutake}, \citenamefont {Tatsumi},\
  and\ \citenamefont {Zhang}}]{Xia2021_PRC103-055812}%
  \BibitemOpen
  \bibfield  {author} {\bibinfo {author} {\bibfnamefont {C.-J.}\ \bibnamefont
  {Xia}}, \bibinfo {author} {\bibfnamefont {T.}~\bibnamefont {Maruyama}},
  \bibinfo {author} {\bibfnamefont {N.}~\bibnamefont {Yasutake}}, \bibinfo
  {author} {\bibfnamefont {T.}~\bibnamefont {Tatsumi}}, \ and\ \bibinfo
  {author} {\bibfnamefont {Y.-X.}\ \bibnamefont {Zhang}},\ }\href {\doibase
  10.1103/PhysRevC.103.055812} {\bibfield  {journal} {\bibinfo  {journal}
  {Phys. Rev. C}\ }\textbf {\bibinfo {volume} {103}},\ \bibinfo {pages}
  {055812} (\bibinfo {year} {2021})}\BibitemShut {NoStop}%
\bibitem [{\citenamefont {Maruyama}\ \emph {et~al.}(2005)\citenamefont
  {Maruyama}, \citenamefont {Tatsumi}, \citenamefont {Voskresensky},
  \citenamefont {Tanigawa},\ and\ \citenamefont
  {Chiba}}]{Maruyama2005_PRC72-015802}%
  \BibitemOpen
  \bibfield  {author} {\bibinfo {author} {\bibfnamefont {T.}~\bibnamefont
  {Maruyama}}, \bibinfo {author} {\bibfnamefont {T.}~\bibnamefont {Tatsumi}},
  \bibinfo {author} {\bibfnamefont {D.~N.}\ \bibnamefont {Voskresensky}},
  \bibinfo {author} {\bibfnamefont {T.}~\bibnamefont {Tanigawa}}, \ and\
  \bibinfo {author} {\bibfnamefont {S.}~\bibnamefont {Chiba}},\ }\href
  {\doibase 10.1103/PhysRevC.72.015802} {\bibfield  {journal} {\bibinfo
  {journal} {Phys. Rev. C}\ }\textbf {\bibinfo {volume} {72}},\ \bibinfo
  {pages} {015802} (\bibinfo {year} {2005})}\BibitemShut {NoStop}%
\bibitem [{\citenamefont {Huang}\ \emph {et~al.}(2021)\citenamefont {Huang},
  \citenamefont {Wang}, \citenamefont {Kondev}, \citenamefont {Audi},\ and\
  \citenamefont {Naimi}}]{Huang2021_CPC45-30002}%
  \BibitemOpen
  \bibfield  {author} {\bibinfo {author} {\bibfnamefont {W.}~\bibnamefont
  {Huang}}, \bibinfo {author} {\bibfnamefont {M.}~\bibnamefont {Wang}},
  \bibinfo {author} {\bibfnamefont {F.}~\bibnamefont {Kondev}}, \bibinfo
  {author} {\bibfnamefont {G.}~\bibnamefont {Audi}}, \ and\ \bibinfo {author}
  {\bibfnamefont {S.}~\bibnamefont {Naimi}},\ }\href {\doibase
  10.1088/1674-1137/abddb0} {\bibfield  {journal} {\bibinfo  {journal} {Chin.
  Phys. C}\ }\textbf {\bibinfo {volume} {45}},\ \bibinfo {pages} {030002}
  (\bibinfo {year} {2021})}\BibitemShut {NoStop}%
\bibitem [{\citenamefont {Wang}\ \emph {et~al.}(2021)\citenamefont {Wang},
  \citenamefont {Huang}, \citenamefont {Kondev}, \citenamefont {Audi},\ and\
  \citenamefont {Naimi}}]{Wang2021_CPC45-030003}%
  \BibitemOpen
  \bibfield  {author} {\bibinfo {author} {\bibfnamefont {M.}~\bibnamefont
  {Wang}}, \bibinfo {author} {\bibfnamefont {W.}~\bibnamefont {Huang}},
  \bibinfo {author} {\bibfnamefont {F.}~\bibnamefont {Kondev}}, \bibinfo
  {author} {\bibfnamefont {G.}~\bibnamefont {Audi}}, \ and\ \bibinfo {author}
  {\bibfnamefont {S.}~\bibnamefont {Naimi}},\ }\href {\doibase
  10.1088/1674-1137/abddaf} {\bibfield  {journal} {\bibinfo  {journal} {Chin.
  Phys. C}\ }\textbf {\bibinfo {volume} {45}},\ \bibinfo {pages} {030003}
  (\bibinfo {year} {2021})}\BibitemShut {NoStop}%
\bibitem [{\citenamefont {Angeli}\ and\ \citenamefont
  {Marinova}(2013)}]{Angeli2013_ADNDT99-69}%
  \BibitemOpen
  \bibfield  {author} {\bibinfo {author} {\bibfnamefont {I.}~\bibnamefont
  {Angeli}}\ and\ \bibinfo {author} {\bibfnamefont {K.}~\bibnamefont
  {Marinova}},\ }\href {\doibase https://doi.org/10.1016/j.adt.2011.12.006}
  {\bibfield  {journal} {\bibinfo  {journal} {At. Data Nucl. Data Tables}\
  }\textbf {\bibinfo {volume} {99}},\ \bibinfo {pages} {69} (\bibinfo {year}
  {2013})}\BibitemShut {NoStop}%
\end{thebibliography}

%

\end{document}